\long\def\comment#1{}
\documentclass[11pt, reqno]{amsart}
\usepackage{harvard}
\usepackage{graphicx,amsmath,amssymb,epsfig}
\usepackage{harvard}
\long\def\comment#1{}
\newtheorem{theorem}{Theorem}

\newtheorem{lemma}{Lemma}

\newtheorem{assumption}{C\!\!}
\theoremstyle{definition}

\newtheorem{remark}{Comment}[section]

\newcommand{\citen}{\citeasnoun}

\newcommand{\be}{\begin{eqnarray}}
\newcommand{\ee}{\end{eqnarray}}

\newcommand{\NN}{\mathbf{N}}

\newcommand{\XX}{\mathcal{X}}

\newcommand{\ind}{ \to_d}
\newcommand{\inp}{ \to_p}
\newcommand{\ba}{\begin{array}}
\newcommand{\ea}{\end{array}}
\newcommand{\bs}{\begin{align}\begin{split}\nonumber}
\newcommand{\bsnumber}{\begin{align}\begin{split}}
\newcommand{\es}{\end{split}\end{align}}

\newcommand{\sss}{\scriptscriptstyle}

\renewcommand{\(}{\left(}
\renewcommand{\)}{\right)}
\renewcommand{\[}{\left[}
\renewcommand{\]}{\right]}
\renewcommand{\hat}{\widehat}

\newcommand{\aT}{A_T}

\newcommand{\QQ}{\widetilde{Q}}

\begin{document}

\pagestyle{empty}

\title[Inference for Extremal Quantile Regression]{
Inference for Extremal Conditional Quantile Models, with an Application
to Market and Birthweight Risks \\
}
\author[ ] {Victor
Chernozhukov$^\dag$  \ \ Iv\'an Fern\'andez-Val$^\S$}

\thanks{First version: May, 2002. This version: \today.  This is a
revision of the paper with the original title ``Extreme Value
Inference for Quantile Regression.'' We would like to thank very
much Takeshi Amemiya, Brigham Fradsen, Hide Ichimura, Jerry Hausman, Peter Hinrichs,
Keith Knight, Roger Koenker, Joseph Romano, the editor Bernard
Salanie, and two referees.}

\thanks{$\dag$  Massachusetts Institute of Technology, Department of
Economics and Operations Research Center, University College London,
CEMMAP. E-mail: vchern@mit.edu.}

\thanks{$\S$ Boston University, Department of Economics. E-mail:
ivanf@bu.edu.}

\begin{abstract}
Quantile regression is an increasingly important empirical tool in
economics and other sciences for analyzing the impact of a set of
regressors on the conditional distribution of an outcome. Extremal
quantile regression, or quantile regression applied to the tails, is
of interest in many economic and financial applications, such as
conditional value-at-risk, production efficiency, and adjustment bands
in (S,s) models.  In this paper we provide feasible inference tools for
extremal conditional quantile models that rely upon extreme value
approximations to the distribution of self-normalized quantile
regression statistics. The methods are simple to implement and can
be of independent interest even in the non-regression case. We
illustrate the results with two empirical examples analyzing extreme fluctuations of a stock return
and extremely low percentiles of live infants'
birthweights in the range between 250 and 1500 grams.\\



\noindent {\sc Monte-Carlo programs and software are available at www.mit.edu/vchern} \\

\end{abstract}

 \maketitle
\pagestyle{plain}\thispagestyle{empty}
\newpage\pagestyle{headings}\setcounter{page}{1}

\section{Introduction and Motivation}

Quantile regression (QR) is an increasingly important empirical tool
in economics and other sciences for analyzing the impact of a set of
regressors $X$  on features of the conditional distribution of an
outcome $Y$ (see Koenker, 2005).   In many applications the features
of interest are the  extremal or tail quantiles of the conditional
distribution.   This paper provides practical tools for performing
inference on these features using extremal QR and
extreme value theory.  The key problem we address is that
conventional inference methods for QR, based on the
normal distribution, are not valid for extremal QR. By using extreme
value theory, which specifically accounts for the extreme nature of
the tail data, we are able to provide inference methods that are
 valid for extremal QR.

Before describing the contributions of this paper in more detail, we
first motivate the use of extremal quantile regression
in specific economic applications.  Extremal
quantile regression provides a useful description of important
features of the data in these applications, generating both
reduced-form facts as well as inputs into estimation of structural
models. In what follows,  $Q_{Y}(\tau|X)$ denotes the conditional
$\tau$-quantile of $Y$ given $X$; extremal conditional quantile
refers to the conditional quantile function $Q_{Y}(\tau|X)$ with the
quantile index $\tau = \epsilon$ or $1-\epsilon$, where $\epsilon$
is close to zero; and extremal quantile regression refers to the
quantile regression estimator of an extremal conditional quantile.

A principal area of economic applications of extremal quantile
regression is  risk management.  One example in this area is
conditional value-at-risk analysis from financial economics
\cite{vl,caviar}.  Here, we are interested in the extremal quantile
$Q_{Y}(\epsilon|X)$ of a return $Y$ to a bank's portfolio,
conditional on various predictive variables $X$, such as the return
to the market portfolio and the returns to portfolios of other
related banks and mortgage providers.  Unlike unconditional extremal
quantiles, conditional extremal quantiles are useful for
stress testing and analyzing the impact of adverse systemic events
on the bank's performance. For example, we can analyze the impact of
a large drop in the value of the market portfolio or of an associated
company on the performance of the bank's portfolio.   The results of
this analysis are useful for determining the level of capital that
the bank needs to hold to prevent bankruptcy in unfavorable states
of the world. Another example comes from health economics, where we
are interested in the analysis of socio-economic determinants $X$ of
extreme quantiles of a child's birthweight $Y$ or other health
outcomes. In this example, very low birthweights are connected with
substantial health problems for the child, and thus extremal quantile regression is useful
to identifying which factors can improve these negative health
outcomes.   We shall return to these examples later in the empirical
part of the paper.

Another primary area of economic applications of extremal quantile
regression deals with describing approximate or probabilistic
boundaries of economic outcomes conditional on pertinent factors. A
first example in this area comes from efficiency analysis in the
economics of regulation, where we are interested in the
probabilistic production frontier $Q_{Y}(1-\epsilon|X)$. This
frontier describes the level of production $Y$ attained by the most
productive $(1-\epsilon)\times 100$ percent of firms,
conditional on input factors $X$ \cite{timmer}.  A second
example comes from the analysis of job search in labor economics,
where we are interested in the approximate reservation wage
$Q_{Y}(\epsilon|X)$. This function describes the wage level,
below which the worker accepts a job only with a small probability
 $\epsilon$, conditional on worker characteristics and other
factors $X$ \cite{flinn}.  A third example deals with estimating
$(S,s)$ rules in industrial organization and macroeconomics
\cite{caballero:engel3}.  Recall that the $(S,s)$ rule is an optimal
policy for capital adjustment, in which a firm allows its capital
stock to gradually depreciate to a lower barrier, and  once the
barrier is reached, the firm adjusts its capital stock sharply to an
upper barrier. Therefore, in a given cross-section of firms, the
extremal conditional quantile functions $Q_{Y}(\epsilon|X)$ and
$Q_{Y}(1-\epsilon|X)$  characterize the approximate
adjustment barriers for observed capital stock $Y$, conditional on a
set of observed factors $X$.\footnote{\citeasnoun{caballero:engel3}
study approximate adjustment barriers using distribution
models; obviously quantile models can also be used.}$^,$\footnote{In
the previous examples, we can set $\epsilon$ to 0 to recover the
exact, non-probabilistic, boundaries in the case with no unobserved
heterogeneity and no (even small) outliers in the data. Our
inference methods cover this exact extreme case, but we recommend
avoiding it because it requires very stringent assumptions.}

The two areas of applications described above are either
non-structural or semi-structural. A third principal area of
economic applications of extremal quantile regressions is structural
estimation of economic models. For instance, in procurement auction
models, the key information about structural parameters is contained in the extreme or
near-extreme conditional quantiles of bids given bidder and auction
characteristics (see e.g. \citeasnoun{chernozhukov:hong:denjump} and
\citeasnoun{hirano:porter}).  We then can estimate and test a structural
model based on its ability to accurately reproduce a collection of
extremal conditional quantiles observed in the data. This indirect
inference approach is called the method-of-quantiles
\cite{koenker:book}. We refer the reader to
\citeasnoun{dp:superconsistent} for a detailed example of this
approach in the context of using $k$-sample extreme
quantiles.\footnote{The method-of-quantiles allows us to estimate
structural models both with and without parametric unobserved
heterogeneity. Moreover, the use of near-extreme quantiles instead
of exact-extreme quantiles makes the method more robust to a small
fraction of outliers or neglected unobserved heterogeneity.}

We now describe the contributions of this paper more specifically.
This paper develops feasible and practical inferential methods based
on extreme value (EV) theory for QR, namely, on the limit law theory
for QR developed in Chernozhukov (2005) for cases where the quantile
index $\tau\in (0,1)$ is either low, close to zero, or high, close
to 1.  Without loss of generality we assume the former. By close to
0, we mean that the order of the $\tau$-quantile, $\tau T$, defined
as the product of quantile index $\tau$ with the sample size $T$,
obeys $\tau T \to k < \infty$ as $T \to \infty$. Under this
condition, the conventional normal laws, which are based on the
assumption that $\tau T$ diverges to infinity, fail to hold, and
different EV laws apply instead. These laws approximate
the exact finite sample law of extremal QR better than
the conventional normal laws. In particular, we
find that when the dimension-adjusted order of the $\tau$-quantile,
$\tau T/d$, defined as the ratio of the order of the $\tau$-quantile
to the number of regressors $d$, is not large, less than about $20$ or
$40$, the EV laws may be preferable to the normal law,
whereas the normal laws may become preferable otherwise.
We suggest this simple rule of thumb for choosing between the EV
laws and normal laws, and refer the reader to Section 5 for more
refined suggestions and recommendations.

Figure 1 illustrates the difference between the EV and normal
approximations to the finite sample distribution of the extremal QR estimators.
We plot the quantiles of these approximations against the quantiles of the exact finite
sample distribution of the QR estimator. We consider different
dimension-adjusted orders in a simple model with only one regressor,
$d=1$, and $T=200$. If either the EV law or the normal law were to
coincide with the true law, then their quantiles would fall
exactly on the 45 degree line shown by the solid line. We see from
the plot that when the dimension-adjusted order $\tau T/d$ is $20$
or $40$, the quantiles of the EV law are indeed very close to the 45
degree line, and in fact are much closer to this line than the
quantiles of the normal law. Only for the case when the effective
order $\tau T/d$ becomes $60$, do the quantiles of the EV law and
normal laws become comparably close to the 45 degree line.

\begin{figure}[!http!]\label{shock}
 \psfigdriver{dvips}
 \epsfig{figure=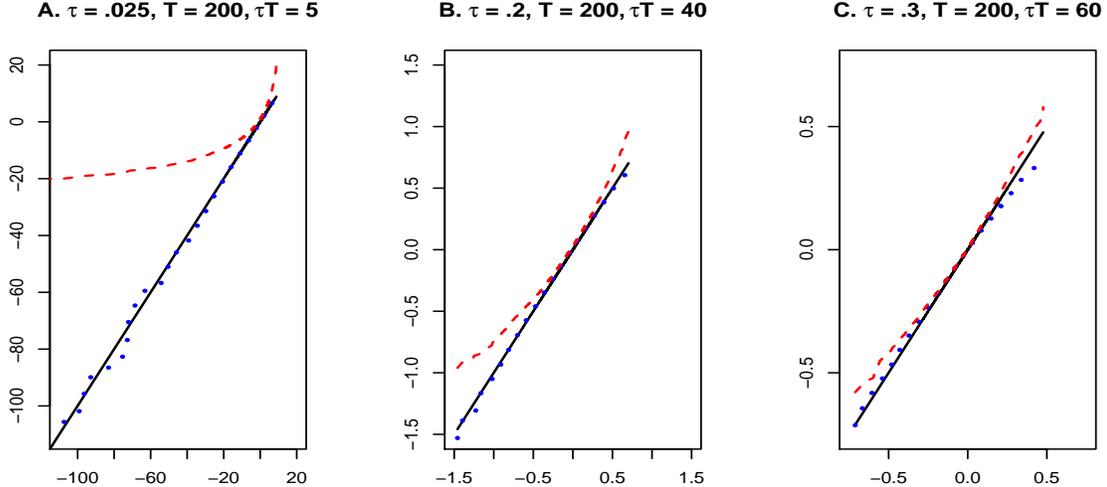, width=6.2in, height=3in}
 \caption{\textbf{Quantiles of the true law of QR  vs. quantiles of EV and normal
 laws}. The figure is based on a simple design with $Y = X + U$, where $U$
 follows a Cauchy distribution and $X=1$.  The solid line ``------" shows the actual quantiles of the true
distribution of QR with quantile index $\tau \in \{ .025, .2, .3\}$.  The dashed line ``- \ - \ -" shows the quantiles
of the  conventional normal law for QR, and the dotted line ``......"  shows the quantiles
of EV law for QR. The figure is based on 10,000 Monte Carlo replications and plots quantiles over the $99\%$ range.
}
\end{figure}

A major problem with implementing the EV approach, at least
in its pure form, is its infeasibility for inference purposes. Indeed, EV approximations rely on canonical
normalizing constants to achieve non-degenerate asymptotic laws.
Consistent estimation of these constants is generally not possible,
at least without making additional strong assumptions. This
difficulty is also encountered in the classical non-regression case;
see, for instance, \citeasnoun{bertailetal} for discussion.
Furthermore, universal inference methods such as the bootstrap
fail due to the nonstandard behavior of extremal QR statistics; see
\citeasnoun{bickel} for a proof in the classical non-regression
case. Conventional subsampling methods with and without replacement
are also inconsistent because the QR statistic diverges in  the
unbounded support case. Moreover, they require consistent estimation
of normalizing constants, which is not feasible in general.

In this paper we develop two types of inference approaches that
overcome all of the difficulties mentioned above:  a resampling approach
 and an analytical approach. We favor the first approach due to its ease of implementation in practice. At the heart of both approaches
 is the use of self-normalized QR (SN-QR) statistics that employ random
normalization factors, instead of generally infeasible normalization
by canonical constants. The use of SN-QR statistics allows us to
derive feasible limit distributions, which underlie either of our
inference approaches. Moreover, our resampling approach is a
suitably modified subsampling method applied to SN-QR statistics.
This approach entirely avoids estimating not only the canonical
normalizing constants, but also all other nuisance tail parameters,
which in practice may be difficult to estimate reliably. Our
construction fruitfully exploits the special relationship between the
rates of convergence/divergence of extremal and intermediate QR statistics,
which allows for a valid estimation of the centering constants
in subsamples.  For completeness we also provide inferential methods
for canonically-normalized QR (CN-QR) statistics,  but we also show
that their feasibility requires much stronger assumptions.

The remainder of the paper is organized as follows. Section 2
describes the model and regularity conditions, and gives an
intuitive overview of the main results. Section 3 establishes the
results that underlie the inferential procedures. Section 4
describes methods for estimating critical values. Section 5
compares inference methods based on EV and normal approximations
through a Monte Carlo experiment. Section 6 presents empirical
examples, and the Appendix collects proofs and all other figures.

\section{The Set Up and Overview of Results}

\subsection{Some Basics} Let a real random variable $Y$ have a continuous
distribution function $F_Y(y) = \Pr[Y \leq y]$. A $\tau$-quantile of
$Y$  is $Q_Y(\tau)= \inf\{y: F_{Y}(y) > \tau\}$ for some $\tau \in
(0,1).$ Let $X$ be a vector of covariates related to $Y$, and
$F_{Y}(y|x)= \Pr[Y \leq y |X=x]$ denote the conditional distribution
function of $Y$ given $X=x$. The conditional $\tau$-quantile of $Y$
given $X=x$ is $Q_Y(\tau | x) = \inf\{y: F_{Y}(y | x)
> \tau\}$ for some $\tau \in (0,1).$  We refer to $Q_{Y}(\tau|x)$, viewed as a function of $x$, as the
$\tau$-quantile regression function.  This function measures the
effect of covariates on outcomes, both at the center and at the
upper and lower tails of the outcome distribution. A conditional
$\tau$-quantile is \textit{extremal} whenever the probability index
$\tau$ is either low or high in a sense that we will make more
precise below. Without loss of generality, we focus the discussion
on low quantiles.

Consider the classical linear functional form for the conditional
quantile function of $Y$ given $X=x$:
 \be\label{star}
Q_Y(\tau|x) = x'\beta(\tau), \quad  \text{ for all } \tau \in
\mathcal{I}=(0, \eta],   \text{ for some } \eta \in (0,1],
 \ee
and for every $x \in \mathbf{X},$ the support of $X$. This linear
functional form is flexible in the sense that it has good
approximation properties. Indeed, given an original regressor
$X^*$, the final set of regressors $X$ can be formed as a vector of
approximating functions. For example, $X$ may include power
functions, splines, and other transformations of $X^*$.

Given a sample of $T$ observations  $\{ Y_t, X_t, t=1,...,T\}$, the
$\tau$-quantile QR estimator $\hat \beta(\tau)$ solves:
\bsnumber\label{canonical1} \hat \beta(\tau) \in \arg \min_{\beta
\in \Bbb{R}^d } \quad \sum_{t=1}^T \rho_{\tau} \( Y_t - X_t' \beta
\),
\end{split}\end{align}
 where $\rho_\tau(u) = (\tau - 1(u<0))u$ is the asymmetric absolute deviation
 function. The median $\tau=1/2$ case
 of  (\ref{canonical1}) was introduced by
 \citeasnoun{laplace:1818} and the general quantile formulation (\ref{canonical1}) by  Koenker and Bassett (1978).

QR coefficients $\hat \beta(\tau)$  can be seen as order statistics
in the regression setting. Accordingly, we will refer to $\tau T$
 as the order of the $\tau$-quantile. A sequence of quantile index-sample size pairs
$\{\tau_T, T\}_{T=1}^\infty$ is said to be an {\it extreme order}
sequence if $\tau_T
 \searrow 0$ and $ \tau_T T \rightarrow k \in (0, \infty)$ as $T \to \infty$;
 an  {\it intermediate order} sequence if $\tau_T
 \searrow 0$ and $ \tau_T T \rightarrow \infty$ as $T \to \infty$;  and a {\it central order} sequence if $\tau_T$
 is fixed as $T \rightarrow \infty$.  Each type of sequence
 leads to different asymptotic approximations to the finite-sample
 distribution of the QR estimator.  The extreme order
 sequence leads to an extreme value (EV) law in large samples, whereas the intermediate and central
 sequences lead to normal laws. As we saw in Figure 1, the EV law
 provides a better approximation to the finite sample law of the QR
 estimator than the normal law.

\subsection{Pareto-type or Regularly Varying Tails}

In order to develop inference theory for extremal QR, we assume
the tails of the conditional distribution of the outcome
variable have Pareto-type behavior, as we formally state in the
next subsection. In this subsection, we recall and discuss the
concept of Pareto-type tails.  The (lower) tail of a distribution
function has Pareto-type behavior if it decays approximately as a
power function, or more formally, a regularly varying function. The
tails of the said form
 are prevalent in economic data, as discovered by V. Pareto in 1895.\footnote{ Pareto called
the tails of this form ``A Distribution Curve for Wealth and
Incomes." Further empirical substantiation has been given by
\citeasnoun{sen}, \citeasnoun{zipf}, \citeasnoun{mandelbrot}, and
\citeasnoun{fama:65}, among others. The mathematical theory of
regular variation in connection to extreme value theory has been
developed by Karamata, Gnedenko, and de Haan.} Pareto-type tails
encompass or approximate a rich variety of tail behavior, including
that of thick-tailed and thin-tailed distributions, having either
bounded or unbounded support.

More formally, consider a random variable $Y$ and define a random
variable $U$ as $U \equiv Y$, if the lower end-point of the support
of $Y$ is $-\infty$, or $U \equiv Y-Q_Y(0)$, if the lower end-point
of the support of $Y$ is $Q_Y(0)>-\infty$.  The quantile function of
$U$, denoted by $Q_U$, then has lower end-point $Q_U(0)= {-\infty}$
or $Q_U(0)=0$. The assumption that the quantile function $Q_U$ and
its distribution function $F_U$ exhibit Pareto-type behavior in
the tails can be formally stated as the following two equivalent
conditions:\footnote{ The notation $a \sim b$ means that $a/b
\rightarrow 1$ as appropriate limits are taken. } \be\label{power 1}
Q_U(\tau) &\sim & L(\tau) \cdot \tau^{-\xi} \ \ \text{ as } \tau
\searrow  0,\label{power 2} \\
F_U(u) &\sim& \bar L(u) \cdot u^{-1/\xi} \ \text{ as } u \searrow
Q_U(0), \ee for some real number $\xi \neq 0$, where $L(\tau)$ is a
nonparametric slowly-varying function at $0$, and $\bar L(u)$ is a
nonparametric slowly-varying function at $Q_U(0)$.\footnote{A
function $u \mapsto L(u)$ is said to be slowly-varying at $s$ if
$\lim_{l \searrow s}[L(l)/L(ml)] = 1$ for any $m>0$.} The leading
examples of slowly-varying functions are the constant function and
the logarithmic function. The number $\xi$ defined in (\ref{power
1}) and (\ref{power 2}) is called the EV index.

The absolute value of $\xi$ measures the heavy-tailedness of the
distribution. A distribution $F_Y$ with Pareto-type tails
necessarily has a finite lower support point if $\xi<0$ and a
infinite lower support point if $\xi>0$. Distributions with $\xi>0$
include stable, Pareto, Student's $t$, and many other distributions.
For example, the $t$ distribution with $\nu$ degrees of freedom has
$\xi=1/\nu$ and exhibits a wide range of tail behavior. In
particular, setting $\nu =1$ yields the Cauchy distribution which
has heavy tails with $\xi=1$, while setting $\nu = 30$ gives a
distribution which has light tails with $\xi=1/30$, and which is
very close to the normal distribution. On the other hand,
distributions with $\xi<0$ include the uniform, exponential,
Weibull, and many other distributions.

It should be mentioned that the case of $\xi=0$ corresponds to the
class of rapidly varying distribution functions.  These
distribution functions have exponentially light tails, with the
normal and exponential distributions being the chief examples. To
simplify the exposition, we do not discuss this case explicitly.
However, since the limit distributions of the main statistics are
continuous in $\xi$, including at $\xi=0$, inference theory for the
case of $\xi=0$ is also included  by taking $\xi \to 0$.

\subsection{The Extremal Conditional Quantile Model and Sampling Conditions}

With these notions in mind, our main assumption is that the response
variable $Y$, transformed by some auxiliary regression line,
$X'\beta_e$, has Pareto-type tails with EV index $\xi$.

\begin{assumption} The conditional quantile function of $Y$ given $X=x$ satisfies equation (\ref{star}) a.s.
Moreover, there exists an auxiliary extremal regression parameter
$\beta_e \in \mathbb{R}^d$, such that the disturbance $V \equiv Y
-X'\beta_e$ has end-point $s=0$ or $s=-\infty$ a.s., and its
conditional quantile function $Q_V(\tau|x)$ satisfies the following
tail-equivalence relationship:
$$Q_V(\tau|x) \sim x'\gamma \cdot Q_U(\tau), \text{ as } \tau
\searrow 0, \text{ uniformly in } x \in \mathbf{X} \subseteq
\mathbb{R}^d,$$ for some quantile function $Q_U(\tau)$ that exhibits
Pareto-type tails with EV index $\xi$ (i.e., it satisfies
(\ref{power 2})), and some vector parameter $\gamma$ such that
$E[X]'\gamma=1$.
\end{assumption}

Since this assumption only affects the tails, it allows covariates
to affect the extremal quantile  and the central quantiles very
differently. Moreover, the local effect of covariates in the tail
is approximately given by $\beta(\tau) \approx \beta_e +  \gamma
Q_U(\tau)$, which allows for a differential impact of
covariates across various extremal quantiles.

\begin{assumption} The conditional quantile density function $\partial Q_V(\tau|x)/\partial \tau$ exists
and satisfies the tail equivalence relationship $\partial
Q_V(\tau|x)/\partial \tau \sim x'\gamma \cdot
\partial Q_U(\tau)/ \partial \tau$ as $\tau \searrow 0$,  uniformly in  $x
\in \mathbf{X},$ where $\partial Q_U(\tau)/
\partial \tau$ exhibits Pareto-type tails as $\tau \searrow 0$ with EV index $\xi+1$.
\end{assumption}
Assumption  C2 strengthens C1 by imposing the existence and
Pareto-type behavior of the conditional quantile density function.
We impose C2 to facilitate the derivation of the main inferential
results.

The following sampling conditions will be imposed.

\begin{assumption} The regressor vector $X=(1,Z')'$ is such that
it has a compact support  $\mathbf{X}$, the matrix $E[XX']$ is
positive definite, and its distribution function $F_X$ satisfies a
non-lattice condition stated in the mathematical appendix (this
condition is satisfied, for instance, when $Z$ is absolutely
continuous).
\end{assumption}

Compactness is  needed to ensure the continuity and robustness of
the mapping from extreme events in $Y$ to the extremal QR
statistics. Even if $X$ is not compact, we can select the data for
which $X$ belongs to a compact region. The non-degeneracy condition
 of $E[XX']$ is standard and guarantees invertibility. The non-lattice
condition is required for the existence of the finite-sample density
of QR coefficients. It is needed even asymptotically because the
asymptotic distribution theory of extremal QR closely resembles the
finite-sample theory for QR, which is not a surprise given the rare
nature of events that have a probability of order $1/T$.

We assume the data are either i.i.d. or weakly dependent.
\begin{assumption} The sequence $\{W_t\}$ with
$W_t =(V_t,X_t)$ and $V_t$ defined in C1, forms a stationary,
strongly mixing process with a geometric mixing rate, that is,  for
some $C>0$
$$\sup_{t} \sup_{A \in \mathcal{A}_{t}, B \in \mathcal{B}_{t+m}} |P(A
\cap B) - P(A) P(B)| \exp(C m) \to 0 \text{ as } m \to \infty,$$
 where $\mathcal{A}_{t} = \sigma(W_{t},
W_{t-1}, ...)$ and $\mathcal{B}_{t} = \sigma(W_{t},W_{t+1}, ...)$.
Moreover, the sequence satisfies a condition that curbs clustering
of extreme events in the following sense: $P(V_t \leq K, V_{t+j}\leq
K|\mathcal{A}_{t}) \leq C P(V_t \leq K|\mathcal{A}_{t})^2$ for all
$K \in [s, \bar K]$, uniformly for all $j \geq 1$ and uniformly for
all $t\geq 1$; here $C>0$ and $\bar{K}>s$ are some constants.
\end{assumption}

A special case of this condition is when the sequence of variables
$\{(V_t, X_t), t \geq 1\}$, or equivalently $\{(Y_t, X_t), t \geq
1\}$, is independent and identically distributed. The assumption of
mixing for  $\{(V_t, X_t), t \geq 1\}$ is standard in econometric
analysis (White 1990), and it is  equivalent to the assumption of
mixing of $\{(Y_t, X_t), t \geq 1\}$. The non-clustering condition
is of the \citeasnoun{meyer}-type and states that the probability of
two extreme events co-occurring at nearby dates is much lower than
the probability of just one extreme event. For example, it assumes
that a large market crash is not likely to be immediately followed
by another large crash. This assumption leads to limit distributions
of QRs as if independent sampling had taken place. The plausibility
of the non-clustering assumption is an empirical matter. We
conjecture that our primary inference method based on subsampling is
valid more generally, under conditions that preserve the rates of
convergence of QR statistics and  ensure existence of their
asymptotic distributions.  Finally we note that the assumptions made
here could be relaxed in certain directions for some of the results
stated below, but we decided to state a single set of sufficient
assumptions for all the results.

\subsection{Overview and Discussion of Inferential Results}

We begin by briefly revisiting the classical non-regression case
to describe some intuition and the key obstacles to
performing feasible inference in our more general regression case.
Then we will describe our main inferential results for the
regression case. It is worth noting that our main inferential
methods, based on self-normalized statistics,  are new and of
independent interest even in the classical non-regression case.

Recall the following classical result  on the
 limit distribution of the extremal sample quantiles $\widehat{Q}_Y(\tau)$ \cite{gnedenko1}:  for any integer $k\geq 1$ and $\tau = k/T$, as $T \to
\infty$,
  \be\label{gnedenko} &  \widehat Z_T(k) =  \aT(\widehat{Q}_Y(\tau)-Q_Y(\tau)) \to_d  \widehat Z_\infty(k) = \Gamma_k^{-\xi} - k^{-\xi},\ee where
 \bsnumber
A_T = 1/ Q_U(1/T), \ \ \ \Gamma_k = \mathcal{E}_1 +...+
\mathcal{E}_k,  \\
 \end{split}\end{align}
and $ (\mathcal{E}_1, \mathcal{E}_2, ... )$ is an independent and
identically distributed sequence of standard exponential variables.
We refer to $\widehat Z_T(k)$ as the canonically normalized (CN)
statistic because it depends on the scaling constant $\aT$. The
variables $\Gamma_k$, entering the definition of the EV
distribution, are {\it gamma} random variables. The limit
distribution of the $k$-th order statistic is therefore a
transformation of a gamma variable.  The EV distribution is not
symmetric and may have significant (median) bias; it has finite
moments if $\xi<0$ and has finite moments of up to order $1/\xi$ if
$\xi >0$. The presence of median bias motivates the use of
median-bias correction techniques, which we discuss in the
regression case below.

Although very powerful, this classical result is not feasible for
purposes of inference on $Q_Y(\tau)$, since the scaling constant
$A_T$ is generally not possible to estimate consistently
\cite{bertailetal}. One way to deal with this problem is to add
strong parametric assumptions on the non-parametric, slowly-varying
function  $L(\cdot)$ in equation (\ref{power 2}) in order to
estimate $A_T$ consistently. For instance, suppose that $Q_{U}(\tau)
\sim L \tau^{-\xi}$. Then one can estimate $\xi$ by the classical
Hill or Pickands estimators,
 and $L$ by $\widehat L =  ( \widehat Q_Y(2 \tau)- \widehat  Q_Y(
\tau)))/ (2^{-\widehat\xi}-1) \tau^{-\widehat \xi})$. We develop the
necessary theoretical results for the regression analog of this
approach, although we will not recommend it as our preferred method.

Our preferred and main proposal to deal with the aforementioned
infeasibility problem is to consider the asymptotics of
the self-normalized (SN) sample quantiles \bsnumber \label{EV
approximation} & Z_T(k) = \mathcal{A}_T (\widehat
Q_Y(\tau)-Q_Y(\tau)) \to_d Z_{\infty}(k)= \frac{\sqrt{k}
(\Gamma_k^{-\xi} - k^{-\xi})}{\Gamma_{ mk }^{-\xi}  - \Gamma_{ k
}^{-\xi}  },\end{split}\end{align} where for $m> 1$ such that $mk$
is an integer,
 \bsnumber\label{SN}
\mathcal{A}_T = \frac{\sqrt{\tau T}}{\widehat Q_Y( m\tau) - \widehat
Q_Y(\tau) } .
 \end{split}\end{align}
Here, the scaling factor $\mathcal{A}_T$ is completely a function of
data and therefore \textit{feasible}.    Moreover, we completely
avoid the need for consistent estimation of $\aT$. This is
convenient because we are not interested in this normalization
constant per se. The limit distribution in (\ref{EV approximation})
only depends on the EV index $\xi$, and its quantiles can be easily
obtained by simulation.  In the regression setting, where the limit
law is a bit more complicated, we develop a form of subsampling to
perform both practical and feasible inference.

Let us now turn to the regression case. Here, we can also consider a
canonically-normalized QR statistic (CN-QR):
 \be  \label{equation: EV limit distribution QR}
 \widehat Z_{T}(k) := \aT \Big ( \hat \beta (\tau) -
\beta(\tau) \Big )  \text{ for }  \aT := 1/ Q_{U}(1/T);\ee and a
self-normalized QR (SN-QR) statistic: \be Z_{T}(k) := \mathcal{A}_T
\Big ( \hat \beta (\tau) - \beta(\tau) \Big ) \text{ for }
\mathcal{A}_T := \frac{\sqrt{\tau T}}{\bar X_T'(
 \hat\beta(m\tau)-\hat\beta(\tau))},
\ee where $\bar X_T = \sum_{t=1}^T X_t/T$ and $m$ is a real number
such that $\tau T (m-1)>d$. The first statistic uses an infeasible
canonical normalization $\aT$, whereas the second  statistic uses a
feasible random normalization.  First, we show that \bsnumber
\label{result sub} \widehat Z_{T}(k) \to_d  \widehat Z_{\infty}(k)
\end{split}\end{align}
where for $\chi=1$ if $\xi < 0$ and $\chi = -1$ if $\xi > 0,$
 \bsnumber \label{result cn} & \widehat Z_{\infty}(k) := \
\chi \cdot \underset{z \in \mathbb{R}^d}{\arg \min } \Bigg [- k
E[X]' (z+k^{-\xi}\gamma) + \ \sum_{t=1}^{\infty} \{ \XX_t'(z
+k^{-\xi}\gamma) - \chi \cdot \Gamma_t^{-\xi}\cdot \XX_t'\gamma
\}_+\ \Bigg]
\end{split}\end{align} where $ \{\Gamma_1, \Gamma_2, ...\} := \{
\mathcal{E}_1, \mathcal{E}_1 + \mathcal{E}_2 ,...\};$
$\{\mathcal{E}_1, \mathcal{E}_2, ... \}$  is an iid sequence of
exponential variables that is independent of $\{\mathcal{X}_1,
\mathcal{X}_2, ...\}$,  an iid sequence with distribution $F_{X}$;
and $\{y\}_+ := \max(0, y)$.
  Furthermore, we show that
 \be\label{result main}  {Z}_{T}(k) \to_d {Z}_{\infty}(k): = \frac{ \sqrt{k} \hat Z_{\infty}(k)}
{E[X]'(\hat Z_{\infty}(m k )-\hat Z_{\infty}(k)) + \chi\cdot
(m^{-\xi}-1)k^{-\xi}}. \ee

The limit laws here are more complicated than in the
non-regression case, but they share some common features. Indeed,
the limit laws depend on the variables $\Gamma_i$ in a crucial way, and
 are not necessarily centered at
zero and can have significant first order median biases. Motivated
by the presence of the first order bias, we develop bias corrections for
the QR statistics in the next section. Moreover, just as in the
non-regression case, the limit distribution of the CN-QR statistic
in (\ref{result cn}) is
 generally infeasible for inference purposes.
  We need to know or estimate the scaling constant $A_T$, which is
  the reciprocal of the extremal quantile  of the variable $U$ defined in C1. That is, we require an estimator $\hat A_T$ such
that $ \hat A_T/A_T \inp 1,$ which is not feasible unless the tail
of $U$ satisfies additional strong parametric restrictions. We
provide additional restrictions below that facilitate estimation of
$A_T$ and hence inference based on CN-QR, although this is not our
preferred inferential method.

Our main and preferred proposal for inference is based on the SN-QR
statistic, which does not depend on $A_T$. We estimate the
distribution of this statistic  using either a variation of
subsampling or an analytical method. A key ingredient here is the
feasible normalizing variable $\mathcal{A}_T$, which is randomly
proportional to the canonical normalization $A_T$, in the sense that
$\mathcal{A}_T/A_T$ is a random variable in the limit.\footnote{The
idea of feasible random normalization has been used in other
contexts (e.g. t-statistics). In extreme value theory,
\citeasnoun{Dekkers:dehaan} applied a similar random normalization
idea to extrapolated quantile estimators of intermediate order
in the non-regression setting, precisely to produce limit
distributions that can be easily used for inference. In time series,
\citeasnoun{kiefer:vogelsang} have used feasible inconsistent
estimates of the variance of asymptotically normal estimators. } An
advantage of the subsampling method over the analytical methods is
that it does not require estimation of the nuisance parameters $\xi$
and $\gamma$.
 Our subsampling approach is different from conventional
subsampling in the use of recentering terms and random
normalization. Conventional subsampling that uses recentering by the
full sample estimate $\widehat \beta(\tau)$ is not consistent when
that estimate is diverging; and here we indeed have $A_T \to 0$ when
$\xi>0$. Instead, we recenter by intermediate order
QR estimates in subsamples, which will diverge at a slow enough speed to
estimate the limit distribution of SN-QR consistently. Thus, our
subsampling approach explores the special relationship between the
rates of convergence/divergence of extremal and intermediate QR
statistics and should be of independent interest even in a
non-regression setting.

This paper contributes to the existing literature by  introducing
general feasible inference methods for extremal quantile regression.
Our inferential methods rely in part  on the limit results in
\citeasnoun{victor:annals}, who derived EV limit laws for CN-QR
under the extreme order condition  $\tau T \to k>0$. This theory,
however, did not lead directly to any feasible, practical inference
procedure. \citeasnoun{feigin}, \citeasnoun{victor:nex}, \citeasnoun{portnoy:jur},
and \citeasnoun{knight:linear}  provide
related limit results for canonically normalized linear programming
estimators where $\tau T \searrow 0$, all in different contexts and
at various levels of generality. These limit results likewise did not
provide feasible inference theory. The linear programming estimator
is well suited to the problem of estimating finite deterministic
boundaries of data, as in image processing and other technometric
applications. In contrast, the current approach of taking $\tau T
\to k>0$ is more suited to econometric applications, where interest
focuses on the ``usual" quantiles located near the minimum or
maximum and where the boundaries may be unlimited. However, some of
our theoretical developments are motivated by and build upon this
previous literature. Some of our proofs rely on the elegant
epi-convergence framework of \citeasnoun{geyer} and
\citeasnoun{knight}.

\section{Inference and Median-Unbiased Estimation Based on Extreme Value Laws}

This section establishes the main results that underlie our
inferential procedures.

\subsection{Extreme Value Laws for CN-QR and SN-QR Statistics}

Here we verify that that the CN-QR statistic
$\widehat Z_{T}(k)$ and SN-QR statistic $Z_T(k)$  converge to the limit variables
${\widehat Z}_{\infty}(k)$ and $Z_{\infty}(k)$, under the condition
that  $\tau T \to k >0$ as $T \to \infty$.

\begin{theorem}[Limit Laws for Extremal SN-QR and CN-QR] Suppose conditions C1, C3 and C4 hold. Then as $\tau T \to k
>0$ and $T \to \infty$,  (1) the SN-QR statistic of order $k$ obeys $$ Z_{T}(k)  \ind {Z}_{\infty}(k),$$
for any $m$ such that $k (m-1)>d$,  and (2) the CN-QR  statistic of order $k$ obeys $$
\widehat Z_{T}(k)  \ind {\widehat Z}_{\infty}(k).$$

\end{theorem}

\begin{remark}  The condition that $k(m-1)>d$ in the definition of SN-QR ensures that $\beta(m\tau) \neq \beta(\tau)$ and therefore the
normalization by $\mathcal{A}_T$ is well defined. This is a
consequence of Theorem 3.2 in \citeasnoun{bassett:1982} and
existence of the conditional density of $Y$ imposed in assumption
C2. Result 1 on SN-QR statistics is the main new result that we will
exploit for inference. Result 2 on CN-QR statistics is needed
primarily for auxiliary purposes. \citeasnoun{victor:annals}
presents some extensions of result 2.
\end{remark}

\begin{remark}  When $Q_Y(0|x)> - \infty$, by C1
$Q_Y(0|x)$ is equal to $x'\beta_e$ and is the conditional lower boundary of $Y$. The proof
of Theorem 1 shows that
$$
A_T (\widehat \beta(\tau) -\beta_e) \to_d  \widetilde Z_{\infty}(k):= \widehat Z_{\infty}(k) - k^{-\xi}
\text{ and }
\mathcal{A}_T (\widehat \beta(\tau) -\beta_e) \to_d  \widetilde Z_{\infty}(k)/( \widetilde Z_{\infty}(mk) -  \widetilde Z_{\infty}(k)).
$$
We can use these results and analytical and subsampling methods presented below to perform median unbiased estimation and inference on the boundary parameter $\beta_e$.\footnote{ To estimate the critical values, we can use either analytical or  subsampling methods presented below, with the difference that in subsampling we need to recenter by the full sample estimate $\widehat \beta_e= \widehat \beta(1/T)$.}
\end{remark}

\subsection{Generic Inference and Median-Unbiased Estimation}

We outline two procedures for conducting inference and constructing
asymptotically median unbiased estimates of linear functions $\psi'
\beta(\tau) $ of the coefficient vector $\beta(\tau)$, for some
non-zero vector $\psi$.

\emph{\textbf{1. Median-Unbiased Estimation and Inference Using
SN-QR.}}  By Theorem 1, $\psi' \mathcal{A}_T (\hat \beta(\tau)-
\beta(\tau)) \ind \psi' Z_{\infty}(k).$ Let $c_{\alpha}$ denote the
$\alpha$-quantile of $\psi' Z_{\infty}(k)$ for $0 < \alpha \leq .5$.
Given $\hat c_{\alpha}$, a consistent estimate of $c_{\alpha}$, we
can construct an asymptotically median-unbiased estimator and a
$(1-\alpha)$\%-confidence interval for $\psi' \beta(\tau)$ as $$
\psi' \hat \beta(\tau) - \hat c_{1/2}/\mathcal{A}_T \ \text{ and } \
[\psi' \hat \beta(\tau) - \hat c_{1-\alpha/2}/\mathcal{A}_T, \psi'
\hat \beta(\tau) - \hat c_{\alpha/2}/\mathcal{A}_T],$$ respectively.
The bias-correction term and the limits of the confidence interval
depend on the random scaling $\mathcal{A}_T$. We provide consistent
estimates of $c_{\alpha}$ in the next section.

\begin{theorem}[Inference and median-unbiased estimation using SN-QR] Under the conditions
of Theorem 1, suppose we have $\hat c_{\alpha}$ such that $\hat
c_{\alpha} \inp c_{\alpha}$.  Then,
$$ \lim_{T \to \infty} P\{ \psi' \hat \beta(\tau) - \hat
c_{1/2}/\mathcal{A}_T \leq \psi' \beta(\tau)\} = 1/2$$ and
$$\lim_{T \to \infty} P\{\psi' \hat \beta(\tau) - \hat
c_{1-\alpha/2}/\mathcal{A}_T \leq \psi' \beta(\tau) \leq  \psi' \hat
\beta(\tau) - \hat c_{\alpha/2}/\mathcal{A}_T\} = 1-\alpha.$$
\end{theorem}

\vspace{.3in }

\emph{\textbf{2. Median Unbiased Estimation and Inference Using
CN-QR.}} By Theorem 1, $
 \psi' A_T (\hat \beta(\tau)- \beta(\tau)) \ind
\psi' \widehat Z_{\infty}(k).$ Let $c_{\alpha}'$ denote the
$\alpha$-quantile of $\psi' \widehat Z_{\infty}(k)$ for $0 < \alpha
\leq .5$. Given $\hat A_T$, a consistent estimate of $A_T$,  and
$\hat c_{\alpha}'$, a consistent estimate of $c_{\alpha}'$, we can
construct an asymptotically median-unbiased estimator and a
$(1-\alpha)$\%-confidence interval for $\psi' \beta(\tau)$ as
$$\psi' \hat \beta(\tau) - \hat c_{1/2}'/\widehat{A}_T \ \text{ and
} \ \ [\psi' \hat \beta(\tau) - \hat c_{1-\alpha/2}'/\widehat{A}_T,
\psi' \hat \beta(\tau) - \hat c_{\alpha/2}'/\widehat{A}_T],$$
respectively.

As mentioned in Section 2, construction of consistent estimates of
$A_T$ requires additional strong restrictions on the underlying
model as well as additional steps in estimation. For example,
suppose the nonparametric slowly varying component $L(\tau)$ of
$A_T$ is replaced by a constant $L$, i.e. suppose that as $\tau
\searrow 0$
 \be\label{restrictive}
1/Q_{U}(\tau) = L \cdot  \tau^{\xi} \cdot (1 + \delta(\tau)) \text{
for some } \  L \in \Bbb{R}, \text{ where } \ \delta(\tau)\to 0.
 \ee
We can estimate the constants $L$ and $\xi$ via Pickands-type
procedures:
 \be\label{tail_estimates}
\hat \xi= \frac{-1}{\ln 2} \ln \frac{{\bar X_T}'(\widehat \beta(4
\tau_T)- \widehat \beta( \tau_T))}{\bar X_T'(\widehat \beta(2
\tau_T)- \widehat \beta(\tau_T))}
 \text{ and }
  \hat L =  \frac{\bar X_T'(\widehat \beta(2 \tau_T)- \widehat \beta( \tau_T))
                }
                {
                (2^{-\hat\xi}-1)\cdot \tau^{-\hat \xi}
                },
 \ee
where  $\tau_T$ is chosen to be of an intermediate order, $\tau_T T
\to \infty$ and $\tau_T \to 0$.  Theorem 4 in Chernozhukov
\citeyear{victor:annals} shows that under C1-C4, condition
(\ref{restrictive}), and additional conditions on the sequence
$(\delta(\tau_T), \tau_T)$,\footnote{ The rate convergence of $\hat
\xi$ is $\max[\frac{1}{\sqrt{\tau_T T}}, \ln \delta(\tau_T)]$, which
gives the following condition on the sequence $(\delta(\tau_T),
\tau_T)$ : $
 \max[\frac{1}{\sqrt{\tau_T T}}, \ln \delta(\tau_T)] = o(1/\ln T)$.}
$\hat \xi = \xi + o(1/\ln T)$ and $\hat L \inp L$, which produces
the required consistent estimate $\hat A_T = \hat L
(1/T)^{-\hat\xi}$ such that $\hat A_T/ A_T \to_p 1.$ These
additional conditions on the tails of $Y$ and on the sequence
$(\delta(\tau_T), \tau_T)$ highlight the drawbacks of this inference
approach relative to the previous one.

We provide consistent estimates of $c_{\alpha}'$ in the next
section.

\begin{theorem}[Inference and Median-Unbiased Estimation using CN-QR]
 Assume the conditions of Theorem 1 hold. Suppose
that we have $\hat A_T$ such that $\hat A_T/A_T \inp 1$ and $\hat
c_{\alpha}'$ such that $\hat c_{\alpha} \inp c_{\alpha}'$. Then,
$$\lim_{T \to \infty}P\{ \psi' \hat \beta(\tau) -\hat c_{1/2}'/\hat
A_T \leq \psi' \beta(\tau)\} = 1/2$$ and $$ \lim_{T \to
\infty}P\{\psi' \hat \beta(\tau) - \hat c_{1-\alpha/2}'/\hat A_T
\leq \psi' \beta(\tau) \leq  \psi' \hat \beta(\tau) - \hat
c_{\alpha/2}/\hat A_T\} = 1-\alpha.$$
\end{theorem}

\section{Estimation of Critical Values}

\subsection{Subsampling-Based Estimation of Critical Values.} Our resampling
method for inference uses subsamples to estimate the distribution of
SN-QR, as in standard subsampling.  However, in contrast to the
subsampling, our method bypasses estimation of the unknown
convergence rate $A_T$ by using self-normalized statistics. Our
method also employs a special recentering that allows us to avoid the inconsistency of standard subsampling due to
diverging QR statistics when $\xi>0$.

The method has the following steps. First, consider all subsets of
the data $\{W_t=(Y_t,X_t),$ $t =1,...,T\}$ of size $b$; if $\{W_t\}$
is a time series, consider $B_T = T -b +1$ subsets of size $b$ of
the form $\{ W_i, ..., W_{i+ b -1}\}$. Then compute the analogs of
the SN-QR statistic, denoted $\widehat V_{i, b}$ and defined below
in equation (\ref{subsample_analog}), for each $i$-th subsample for
$i=1,..., B_{\sss T}$. Second, obtain $ \hat c_\alpha$ as the sample
$\alpha$-quantile of $\{ \widehat V_{i,b,T}, i=1,...,B_T \}$. In
practice, a smaller number $B_T$ of randomly chosen subsets can be
used, provided that $B_T \rightarrow \infty$ as $ T \rightarrow
\infty$. (See Section 2.5 in \citeasnoun{romano:book}.)
\citeasnoun{romano:book} and \citeasnoun{bertailetal} provide rules
for the choice of subsample size $b$.

The SN-QR statistic for the full sample of size $T$ is:
 \be
V_{T} : =   \mathcal{A}_T \psi'( \hat \beta_{\sss T}(\tau_{\sss T})
- \beta(\tau_{\sss T})) \text{ for } \mathcal{A}_T  = \frac{
\sqrt{\tau_{\sss T} T}}{ \bar X_T ' \big (\hat \beta (m \tau_{\sss
T}) - \hat \beta ( \tau_{\sss T})\big )}, \ee where we can set
$m= (d+p)/(\tau_{\sss T} T) +1 = (d+p)/k +1 +o(1)$, where $p\geq 1$ is the spacing
parameter, which we set to $5$.\footnote{Variation of this parameter from $p=2$ to
$p=20$ yielded similar results in our Monte-Carlo experiments.} In this section we write $\tau_{\sss T}$ to
emphasize the theoretical dependence of the quantile of interest
$\tau$ on the sample size. In each $i$-th subsample of size $b$, we
compute the following analog of $V_T$:
 \be\label{subsample_analog}
\widehat V_{i,b,\sss T}  : =   \mathcal{A}_{i, b, \sss T} \psi'(
\hat \beta_{i, b, \sss T}(\tau_{\sss b}) - \hat \beta(\tau_{\sss
b})) \text{ for }
 \mathcal{A}_{i,b, \sss T}  : =  \frac{\sqrt{\tau_{\sss b} b}}{ \bar
 X_{i,b,T}
' \big(\hat \beta_{i,b, \sss T}  (m \tau_{\sss b}) - \hat \beta_{i,
b, \sss T} ( \tau_{\sss b})\big )},
 \ee
where $\hat \beta(\tau)$ is the $\tau$-quantile regression
coefficient computed using the full sample, $\hat \beta_{i,b, \sss
T}  ( \tau)$ is the $\tau$-quantile regression coefficient computed
using the $i$-th subsample,  $\bar X_{i,b,T}$ is the sample mean of
the regressors in the $i$th subsample, and $\tau_b := \(\tau_{\sss
T} T \)/b$.\footnote{ In practice, it is reasonable to use the
following finite-sample adjustment to $\tau_b$:
 $\tau_b = \min[ \(\tau_T T\)/b, .2]$ if $\tau_T <.2$, and
 $\tau_b = \tau_T$ if $\tau_T \geq .2$.
The idea is that $\tau_T$ is judged  to be non-extremal if
$\tau_T>.2$, and the subsampling procedure reverts to central
order inference.  The truncation of $\tau_b$ by .2 is a finite-sample
adjustment that restricts the key statistics $\widehat V_{i,b,T}$ to
be extremal in subsamples. These finite-sample adjustments  do not
affect the asymptotic arguments.} The determination of $\tau_b$ is a
critical decision that sets apart the extremal order approximation
from the central order approximation. In the latter case, one sets
$\tau_b = \tau_{\sss T}$ in subsamples. In the extreme order
approximation, our choice of $\tau_b$ gives the same extreme order
of $\tau_b b$ in the subsample as the order of $\tau_TT$ in the full
sample.

Under the additional parametric assumptions on the tail behavior
stated earlier, we can estimate the quantiles of the limit
distribution of CN-QR using the following procedure: First, create
subsamples $i=1,...,B_T$ as before and compute in each subsample:
$\widetilde V_{i, b, \sss T} : = \widehat A_{ b} \psi'( \hat
\beta_{i, b, \sss T}(\tau_{\sss b}) - \hat \beta(\tau_{\sss b})),$
where $\widehat A_b$ is any consistent estimate of $A_b$. For
example, under the parametric restrictions specified in
(\ref{restrictive}), set $\widehat A_b=\hat L b^{-\hat \xi}$  for
$\hat L$ and $\hat \xi$ specified in (\ref{tail_estimates}). Second,
obtain $ \hat c_\alpha'$ as the $\alpha$-quantile of $\{ \widetilde
V_{i,b,T}, i=1,...,B_T \}$.

The following theorems establish the consistency of $\hat
c_{\alpha}$ and $\hat c'_{\alpha}$:

\begin{theorem}[Critical Values for SN-QR by Resampling]
 Suppose the assumptions of Theorems 1 and 2 hold, $b/T \to 0, b \to \infty,
 T \to \infty$ and $B_T \to \infty.$ Then  $\hat
c_\alpha \to_p c_{\alpha}$.
\end{theorem}

\begin{theorem}[Critical Values for CN-QR by Resampling]  Suppose
 the assumptions of Theorems 1 and 2 hold,  $b/T \to 0, b \to \infty,
 T \to \infty, B_T \to \infty$, and
$\widehat A_b$ is such that $\hat A_b/ A_b \to 1$. Then $\hat
c_\alpha' \to_p c'_{\alpha}$.
\end{theorem}

\begin{remark} Our subsampling method based on CN-QR or SN-QR
produces consistent critical values in the regression case, and
may also be of independent interest in the non-regression case.  Our method
differs from conventional subsampling in several respects.
First, conventional subsampling uses fixed normalizations $A_T$ or their consistent
estimates.  In contrast, in the case of SN-QR we use the random normalization $\mathcal{A}_T$,
thus avoiding estimation of $A_T$. Second, conventional subsampling recenters by the
full sample estimate $\hat \beta(\tau_T)$. Recentering in this way requires $A_b / A_T \to 0$ \
for obtaining consistency (see Theorem 2.2.1 in \citeasnoun{romano:book}),
but here we have $A_b / A_T \to \infty$ when $\xi >0$. Thus,  when $\xi>0$
the extreme order QR statistics $\hat \beta(\tau_T)$ diverge
when $\xi>0$, and the conventional subsampling is inconsistent. In contrast,
to overcome the inconsistency, our approach instead uses $\hat
\beta(\tau_b)$ for recentering. This statistic itself may diverge,
but because it is an intermediate order QR statistic, the speed of its divergence is strictly slower than that of
$A_T$.  Hence our method of recentering exploits the special structure of order
statistics in both the regression and non-regression cases.
\end{remark}

\subsection{Analytical Estimation of Critical Values} Analytical inference
uses  the  quantiles of the limit distributions found in Theorem 1.
This approach is much more demanding in practice than the previous
subsampling method.\footnote{The  method developed below is also of
independent interest in situations where the limit distributions
involve Poissson processes with unknown nuisance parameters, as, for
example, in Chernozhukov and Hong (2004).}

Define the following random vector: \be\label{define draw} \widehat
Z^*_{\infty}(k) = \hat{\chi} \cdot \underset{z \in
\mathbb{R}^d}{\arg \min } \Bigg [- k \bar{X}_T'
(z+k^{-\hat{\xi}}\hat{\gamma}) + \ \sum_{t=1}^{\infty} \{ \XX_t'(z
+k^{-\hat{\xi}}\hat{\gamma}) - \hat{\chi} \cdot
\Gamma_t^{-\hat{\xi}}\cdot \XX_t'\hat{\gamma}\}_+\ \Bigg], \ee for
some consistent estimates $\hat \xi$ and $\hat \gamma$, e.g., those
given in equation (\ref{equation:tail estimators}); where $\hat \chi
= 1$ if $\hat \xi < 0$ and $\hat \chi = -1$ if $\hat \xi > 0$,
$\{\Gamma_1, \Gamma_2, ...\} = \{ \mathcal{E}_1, \mathcal{E}_1 +
\mathcal{E}_2 ,...\}$; $\{\mathcal{E}_1, \mathcal{E}_2, ... \} $ is
an i.i.d. sequence of standard exponential variables;
$\{\mathcal{X}_1, \mathcal{X}_2, ... \}$ is an i.i.d. sequence with
distribution function $\widehat F_{X}$, where $\widehat F_{X}$ is
any smooth consistent estimate of $F_X$, e.g., a smoothed empirical
distribution function of the sample $\{X_i, i
=1,...,T\}$.\footnote{We need smoothness of the distribution
regressors $\XX_i$ to guarantee uniqueness of the solution of the
optimization problem (\ref{define draw}); a similar device is used
by \citeasnoun{hall:lad} in the context of Edgeworth expansion for
median regression. The empirical distribution function (edf) of
$X_i$ is not suited for this purpose, since it assigns point masses
to sample points. However, making random draws from the edf and
adding small noise with variance that is inversely proportional to
the sample size produces draws from a smoothed empirical
distribution function which is uniformly consistent with respect to
$F_X$.} Moreover, the sequence $\{\mathcal{X}_1, \mathcal{X}_2, ...
\}$ is independent from $\{ \mathcal{E}_1, \mathcal{E}_2 ,...\}$.
Also,
 let $ {Z}^*_{\infty}(k) = \sqrt{k} \hat Z^*_{\infty}(k)/ [ \bar
X_T'(\hat Z^*_{\infty}(m k )-\hat Z^*_{\infty}(k)) + \hat
\chi(m^{-\hat \xi}-1)k^{-\hat \xi} ]$. The estimates $\hat
c_{\alpha}'$ and $\hat c_{\alpha}$ are obtained by taking
$\alpha$-quantiles of the variables $\psi' \widehat Z^*_{\infty}(k)$
and $\psi'Z^*_{\infty}(k)$, respectively. In practice, these
quantiles can only be evaluated numerically as described below.

The analytical inference procedure requires consistent estimators of
$\xi$ and $\gamma$.  Theorem 4.5 of \citeasnoun{victor:annals}
provides the following estimators based on Pickands-type procedures:
\be \label{equation:tail estimators}
 \hat \xi= \frac{-1}{\ln 2} \ln
\frac{{\bar X_T}'(\widehat \beta(4 \tau_T)- \widehat \beta(
\tau_T))}{{\bar X_T}'(\widehat \beta(2\tau_T)- \widehat
\beta(\tau_T))}  \text{ and } \hat \gamma= \frac{\widehat \beta(2
\tau_T)- \widehat \beta( \tau_T)}{{\bar X_T}'(\widehat \beta(2
\tau_T)- \widehat \beta(\tau_T))}, \ee which is consistent if
$\tau_T T \to \infty$ and $\tau_T \to 0$.

\begin{theorem}[Critical Values for SN-QR by Analytical Method] Assume the conditions
of Theorem 1 hold. Then for any estimators of the nuisance
parameters such that $\hat \xi \to_p \xi$ and $\hat \gamma \to_p
\gamma$, we have that $\hat c_{\alpha} \to_p c_{\alpha}.$
\end{theorem}

\begin{theorem}[Critical Values for CN-QR by Analytical Method] Assume
the conditions of Theorem 1 hold. Then, for any estimators of the
nuisance parameters such that $\hat \xi \to_p \xi$ and $\hat \gamma
\to_p \gamma$, we have that $\hat c_{\alpha}' \to_p c_{\alpha}'.$
\end{theorem}


\begin{remark}
Since the distributions of $\widehat Z_{\infty}(k)$ and
$Z_{\infty}(k)$ do not have closed form, except in very special
cases, $\hat c_{\alpha}'$ and $\hat c_{\alpha}$ can be obtained
numerically via the following Monte Carlo procedure. First, for each
$i=1,...,B$ compute $\widehat Z^*_{i,\infty}(k)$ and
$Z^*_{i,\infty}(k)$ using formula (\ref{define draw}) by simulation,
where the infinite summation is truncated at some finite value $M$.
Second, take $\hat c'_{\alpha}$ and $\hat c_{\alpha}$ as the sample
$\alpha$-quantiles of the samples $\{\psi'\widehat
Z^*_{i,\infty}(k), i=1,...,B\}$ and $\{\psi' Z^*_{i,\infty}(k),
i=1,...,B\}$, respectively. We have found in numerical experiments
that choosing $M \geq 200$ and $B\geq 100$ provides accurate
estimates.
\end{remark}

\section{Extreme Value  vs. Normal Inference: Comparisons}

\subsection{Properties of Confidence Intervals with Unknown Nuisance Parameters}
In this
section we compare the inferential performance of normal and
extremal confidence intervals (CI) using the model: $Y_t =X_t'\beta
+ U_t$, $t=1,...,500$, $d=7$, $\beta_j=1$ for $j \in \{1,...,7\}$,
where the disturbances $\{U_t\}$ are i.i.d. and follow either
 (1)
a $t$ distribution with $\nu \in \{1,3,30\}$ degrees of freedom, or
(2)  a Weibull distribution with the shape parameter $\alpha \in \{
1, 3, 30 \}$.  These distributions have EV indexes $\xi=1/\nu \in
\{1, 1/3, 1/30\}$ and $\xi = -1/\alpha \in \{-1, -1/3, -1/30\}$,
respectively. Regressors are drawn with replacement from the
empirical application in Section 6.1 in order to match a real
situation as closely as possible.\footnote{These data as well as the
Monte-Carlo programs are deposited at www.mit.edu/vchern.} The
design of the first type corresponds to tail properties of financial
data, including returns and trade volumes; and the design of the
second type corresponds to tail properties of microeconomic data,
including birthweights, wages, and bids. Figures 2 and 3 plot
coverage properties of CIs for the intercept and one of the slope
coefficients based on subsampling the SN-QR statistic with $B_T =
200$ and $b = 100$, and on the normal inference method suggested by
Powell (1986) with a Hall-Sheather type rule for the bandwidth
suggested in \citen{koenker:book}.\footnote{ The alternative options
implemented in the statistical package R to obtain standard errors
for the normal method give similar results. These results are available from the authors upon request.} The figures are based on QR estimates
for $\tau \in \{.01, .05, .10, .25, .50 \}$, i.e. $\tau T \in \{5,
25, 50, 125, 250 \}$.

When the disturbances follow $t$ distributions, the extremal CIs have
good coverage properties, whereas the normal CIs typically
\emph{undercover} their performance deteriorates in the
degree of heavy-tailedness and improves in the index $\tau T$.
 In heavy-tailed cases ($\xi \in\{1,1/3\}$) the normal CIs
  substantially undercover for extreme quantiles, as might be expected
  from the fact that the normal distribution fails to capture
  the heavy tails of the actual distribution of the QR statistic.
In the thin-tailed case ($\xi=1/30$), the normal CIs still undercover
for extreme quantiles. The extremal CIs perform consistently better
than normal CIs, giving coverages close to the nominal level of
$90\%$.

When the disturbances follow Weibull distributions, extremal CIs continue
to have good coverage properties, whereas normal CIs either
\emph{undercover} or \emph{overcover}, and their performance
deteriorates in the degree of heavy-tailedness and improves in the
index $\tau T$.  In heavy-tailed cases ($\xi = -1$) the normal CIs
strongly overcover, which results from the
  overdispersion of the normal distribution relative to the actual
  distribution of QR statistics. In the thin-tailed cases
  ($\xi=-1/30$) the normal CIs undercover and their performance improves in the index $\tau
  T$. In all cases, extremal CIs perform better than normal CIs, giving coverage
rates  close to the nominal level of $90\%$ even for central
quantiles.

We also compare forecasting properties of ordinary QR estimators and
median-bias-corrected QR estimators of the intercept and slope
coefficients, using  the median absolute deviation and median bias
as measures of performance (other measures may not be well-defined).
We find that the gains to bias-correcting appear to be very small,
except in the finite-support case with disturbances that are
heavy-tailed near the boundary. We do not report these results for
the sake of brevity.

\subsection{Practicalities and Rules of Thumb}
Equipped with both simulation experiments and practical experience,
we provide a simple rule-of-thumb for the application of
extremal inference. Recall that the order of a sample
$\tau$-quantile in the sample of size $T$ is the number $\tau T$
(rounded to the next integer). This order plays a crucial role in
determining whether extremal inference or central inference should
be applied. Indeed, the former requires $\tau T \to k$ whereas the
latter requires $\tau T \to \infty$. In the regression case, in
addition to the number $\tau T$, we need to take into account the
number of regressors. As an example, let us consider the case where
all $d$ regressors are indicators that equally divide the sample of
size $T$ into subsamples of size $T/d$. Then the QR statistic will
be determined by sample quantiles of order $\tau T/d$ in each of
these $d$ subsamples. We may therefore think of the number $\tau
T/d$ as being a dimension-adjusted order for QR. A common simple
rule for the application of the normal law is that the sample size
is greater than 30. This suggests we should use extremal
inference whenever $\tau T/d \lesssim 30$. This simple rule may or
may not be conservative. For example, when regressors are
continuous, our computational experiments indicate that normal
inference performs as well as extremal inference as soon as $\tau T/d
\gtrsim 15-20$, which suggests using extremal inference when $\tau
T/d \lesssim 15-20$ for this case. On the other hand, if we have an
indicator variable that picks out $2\%$
 of the entire sample, as in the birthweight application presented below,
then the number of observations below the fitted quantile for this
subsample will be $\tau T/50$, which motivates using extremal
inference when $\tau T/50 \lesssim 15-20$ for this case. This rule
is far more conservative than the original simple rule. Overall, it
seems prudent to use both extremal and normal inference methods in
most cases, with the idea that the discrepancies between the two can
indicate extreme situations. Indeed, note that our methods based on
subsampling perform very well even in the non-extreme cases (see
Figures 2 and 3).

\section{Empirical Examples}

\subsection{Extremal Risk of a Stock}

We consider the problem of finding factors that affect the
value-at-risk of the Occidental Petroleum daily stock return, a
problem that is interesting for both economic analysis and
real-world risk management.\footnote{ See \citeasnoun{hahn},
\citeasnoun{diebold:evt}, \citeasnoun{vl}, and
 \citeasnoun{caviar}.}
Our data set consists of 1,000 daily observations covering the period
1996-1998. The dependent variable $Y_t$ is the daily return of the
Occidental Petroleum stock and the regressors $X_{1t}$, $X_{2t}$,
and $X_{3t}$ are the lagged return on the spot price of oil, the
lagged one-day return of the Dow Jones Industrials index (market
return), and the lagged own return $Y_{t-1}$, respectively. We use a
flexible asymmetric linear specification where $X_t = (1, X_{1t}^+,
X_{1t}^-, X_{2t}^+, X_{2t}^-,X_{3t}^+, X_{3t}^-)$ with $X_{jt}^+ =
\max(X_{jt},0)$, $X_{jt}^- = -\min(X_{jt},0)$ and $j \in \{1,2,3\}$.

We begin by stating overall estimation results for the basic
predictive linear model. A detailed specification and
goodness-of-fit analysis of this model has been given in
\citeasnoun{vl}, whereas here we focus on the extremal analysis in
order to illustrate the new inferential tools. Figure 4 plots QR
estimates $\hat \beta(\tau) = ( \hat \beta_j(\tau), j =0,...,7)$
along with 90\% pointwise confidence intervals. We use both extremal
CIs (solid lines) and normal CIs (dashed lines). Figures 5 and 6 plot
bias-corrected QR estimates along with pointwise CIs for the lower
and upper tails, respectively.

We focus the discussion on the impact of downward movements of the
explanatory variables, namely $X_{1t}^-$, $X_{2t}^-$, and
$X_{3t}^-$, on the  extreme risk, that is, on the  low
conditional/predicted quantiles of the stock return. The estimate of
the coefficient on the negative spot price of oil, $X_{1t}^-$, is
positive in the lower tail of the distribution and negative in the
center, but it is not statistically significant at the $90\%$ level.
However, the extremal CIs indicate that the distribution of the QR
statistic is asymmetric in the far left tail, hence the economic
effect of the spot price of oil may potentially be quite strong.
Thus, past drops in the spot price of oil potentially strongly
decrease the extreme risk.  The estimate of the coefficient on the
negative market return, $X_{2t}^-$, is significantly negative in the
far left tail but not in the center of the conditional distribution.
From this we may conclude that the past market drops appear to
significantly increase the extreme risk.  The estimates of the
coefficient on the negative lagged own return, $X_{3t}^-$ are
significantly negative in the lower half of the conditional
distribution. We may conclude that past drops in own return
significantly increase extreme and intermediate risks.

Finally, we compare the CIs produced by extremal inference and normal
inference. This empirical example closely matches the Monte-Carlo
experiment in the previous section with heavy-tailed $t(3)$
disturbances. From this experiment, we expect that in the empirical
example normal CIs would understate the estimation uncertainty and
would be considerably more narrow than extremal CIs in the tails. As
shown in Figures 5 and 6, normal CIs are indeed much more narrow than
extremal CIs at $\tau < .15$ and $\tau > .85$.

\subsection{Extremal Birthweights} We investigate the impact of various demographic
characteristics and maternal behavior on extremely low quantiles of
birthweights of live infants born in the United States to black
mothers of ages between 18 and 45. We use the June 1997 Detailed
Natality Data published by the National Center for Health
Statistics. Previous studies by \citeasnoun{abreveya} and
\citeasnoun{koenker:hallock} used the same data set, but they
focused the analysis on typical birthweights, in a range between
2000 and 4500 grams. In contrast, equipped with extremal inference,
we now venture far into the tails and study extremely low
birthweight quantiles, in the range between 250 and 1500 grams. Some
of our findings  differ sharply from previous results for typical
non-extremal quantiles.

Our decision to focus the analysis on black mothers is motivated by
Figure 7 which shows a troubling heavy tail of low birthweigts for
black mothers. We choose a linear specification similar to Koenker
and Hallock (2001).   The response variable is the birthweight
recorded in grams. The set of covariates include: `Boy,' an
indicator of infant gender; `Married,' an indicator of whether the
mother was married or not; `No Prenatal,' `Prenatal Second,' and
`Prenatal Third,' indicator variables that divide the sample into 4
categories: mothers with no prenatal visit (less than $1\%$ of the
sample), mothers whose first prenatal visit was in the second
trimester, and mothers whose first prenatal visit was in the third
trimester (The baseline category is mothers with a first visit in
the first trimester, which constitute $83\%$ of the sample);
`Smoker,'  an indicator of whether the mother smoked during
pregnancy; `Cigarettes/Day,' the mother's reported average number of
cigarettes smoked per day;
`Education,' a categorical variable taking a value of $0$ if the
mother had less than a high-school education, $1$ if she completed
high school education, $2$ if she obtained some college education,
and $3$ if she graduated from college; `Age' and `Age$^2$,' the
mother's age and the mother's age squared, both in deviations from
their sample means.\footnote{We exclude variables related to
mother's weight gain during pregnancy because they might be
simultaneously determined with the birth-weights.} Thus the control
group consists of mothers of average age who had their first
prenatal visit during the first trimester, that have not completed
high school, and who did not smoke.
The intercept in the estimated quantile
regression model will measure quantiles for this group, and will
therefore be referred to as the centercept.

Figures 8 and 9 report estimation results for extremal low quantiles
and typical quantiles, respectively. These figures show point
estimates, extremal 90\% CIs, and normal 90\% CIs. Note that the
centercept in Figure 8 varies from 250 to about 1500 grams,
indicating the approximate range of birthweights that our extremal
analysis applies to. In what follows, we focus the discussion only
on key covariates and on differences between extremal and central
inference.

While the density of birthweights, shown in Figure 7, has a finite lower
support point, it  has little probability mass near the boundary.
This points towards a situation similar to the Monte Carlo
design with Weibull disturbances, where differences between
central and extremal inference occur only sufficiently far in the
tails. This is what we observe in this empirical example as well. For
the most part, normal CIs tend to be at most $15$ percent narrower
than extremal CIs, with the exception of the coefficient on `No
Prenatal', for which normal CIs are twice as narrow as extremal
CIs. Since only $1.9$ percent of mothers had no prenatal care, the
sample size used to estimate this coefficient is only 635, which
suggests that the discrepancies between extremal CIs and central CIs
for the coefficient on `No prenatal' should occur only when $\tau
\lesssim 30/635 = 5 \%$. As Figure 9 shows, differences between
extremal CIs and normal CIs arise mostly when $\tau \lesssim 10\%$.

The analysis of extremal birthweights, shown in Figure 8, reveals
several departures from findings for typical
birthweights in Figure 9. Most surprisingly, smoking appears to have
no negative impact on extremal quantiles, whereas it has  a strong
negative effect on the typical quantiles. The lack of statistical
significance in the tails could be due to selection, where only
mothers confident of good outcomes smoke, or to smoking having
little or no causal effect on very extreme outcomes. This finding
motivates further analysis, possibly using data sets that enable
instrumental variables strategies.

Prenatal medical care has a strong impact on  extremal quantiles and
relatively little impact on typical quantiles, especially in the
middle of the distribution. In particular, the impacts of `Prenatal
Second' and `Prenatal Third' in the tails are very strongly
positive. These effects could be due to mothers confident of good
outcomes choosing to have a late first prenatal visit.
Alternatively, these effects could be due to a late first prenatal
visit providing better means for improving birthweight outcomes. The
extremal CIs for `No-prenatal' includes values between $0$ and $-800$
grams, suggesting that the effect of `No-prenatal' in the tails is
definitely non-positive and may be strongly negative.




\appendix

\pagestyle{headings}
\small

\section{Proof of Theorem 1}

 The proof will be given for
the case when $\xi <0$. The case with $\xi>0$ follows very
similarly.

\textsc{Step 1.} Recall that $V_t = Y_t - X_t'\beta_e$ and consider
the point process $\hat \NN$ defined by $ \hat \NN (F) :=
\sum_{t=1}^T 1 \{ \(\aT V_t, X_t \) \in F\}$ for Borel subsets $F$
of $E := [  0, +\infty
 ) \times \mathbf{X}$.
The point process $\hat \NN$ converges in law in the metric space of
point measure $M_p(E)$, that is equipped with the metric induced by
the topology of vague convergence. The limit process is a Poisson
point process $\NN$ characterized by the mean intensity measure $
m_{\mathbf{N}}(F):= \int_{F} (x'\gamma)^{1/\xi} u^{-1/\xi} d u d
F_X(x).$ Given this form of the mean intensity measure we can
represent \be \label{eq: limit N}
 \NN (F) := \sum_{t=1}^\infty 1 \{ \(J_t, \XX_t \) \in F\}
\ee for all Borel subsets $F$ of $E := [  0, +\infty
 ) \times \mathbf{X}$,
where $J_t = (\XX_t'\gamma) \cdot \Gamma^{-\xi}_t$,  $\Gamma_t =
\mathcal{E}_1 +...+ \mathcal{E}_t,$ for $t \geq 1$,
$\{\mathcal{E}_t, t \geq 1\}$ is an i.i.d. sequence of standard
exponential variables, $\{\XX_t, t \geq 1\}$ is an i.i.d. sequence
from the distribution $F_X$. Note that when $\xi>0$ the same result
and representation holds, except that we define
 $J_i = -(\XX_i'\gamma) \cdot \Gamma^{-\xi}_i$ (with a change of sign).

The convergence in law $\hat \NN \Rightarrow \NN$ follows from the
following steps. First, for any set $F$ defined as intersection of a
bounded rectangle with $E$, we have
 (a) $\lim_{T \to \infty} E \hat \NN(F) = m_{\mathbf{N}}(F)$, which follows from the regular
variation property of $F_U$ and C1, and (b)  $\lim_{T \to \infty} P[
\hat \NN(F)=0] = e^{-m_{\mathbf{N}}(F) }$,  which  follows by
Meyer's \citeyear{meyer} theorem by the geometric strong mixing and
by observing that $ T \sum_{j=2}^{\lfloor T/k\rfloor} P ( (\aT V_1,
X_1) \in F, (\aT V_j, X_j) \in F) \leq O( T \lfloor T/k\rfloor  P
((\aT V_1, X_1) \in F)^2) = O (1/k) $  by C1 and C5. Consequently,
(a) and (b) imply by Kallenberg's theorem \cite{resnick:book} that
$\hat \NN \Rightarrow \NN$, where $\NN$ is a Poisson point process
$\NN$ with intensity measure $m_{\mathbf{N}}$.

\textsc{Step 2.} Observe that $\widetilde Z_{\sss T}(k)$ $:=$ $\aT
(\hat \beta (\tau) - \beta_e)$ $=$ $\arg
 \min_{z \in
\mathbb{R}^d}
 \sum_{t=1}^T
\rho_{ \tau } \big ( \aT V_t - X_t'z \big ) $. To see this define
$z := \aT
 (     \beta - \beta_e    )$.
Rearranging terms gives $ \sum_{t=1}^T \rho_{ \tau } \big ( \aT V_t
- X_t'z \big ) \equiv  - \tau T  \bar X_T'z - \sum_{t=1}^T 1(\aT V_t
\leq X_t z) \big (\aT V_t - X_t'z\big ) + \sum _{t=1}^T \tau \aT
V_t. $ Subtract $\sum _{t=1}^T \tau \aT V_t$ that does not depend on
$z$ and does not affect optimization, and define $$\widetilde
Q_T(z,k) := - \tau T \bar X_T'z  + \sum_{t=1}^T \ell( \aT V_t,
X_t'z) = - \tau T \bar X_T'z+ \int_E \ell(u,x'z) d\widehat
\NN(u,x),$$ where $ \ell(u, v) :=  1(u \leq v)(v - u).$ We have that
$\widetilde Z_{\sss T}(k)$= $\arg
 \min_{z \in \mathbb{R}^d}\widetilde Q_T(z,k)$.

Since $\ell$ is continuous and vanishes outside a compact subset of
$E$, the mapping $N \mapsto \int_E \ell(u,x'z) d  N(u,x)$, which
sends elements $N$ of  the metric space $M_p(E)$, to the real line,
is continuous.  Since $\tau T \bar X_T \to_p k E[X]$ and $\hat \NN
\Rightarrow \NN$, by the Continuous Mapping Theorem we conclude that
the finite-dimensional limit of $z \mapsto \widetilde Q_T(z,k)$ is
given by
$$ z \mapsto \widetilde Q_{\infty} (z, k) :=  - k E[X]'z + \int_{E} \ell(j,
x'z) d \NN (j, x) := - k E[X]'z + \sum_{t=1}^\infty \ell(J_t,
\XX_t'z). $$

Next we recall the Convexity Lemma of
\citeasnoun{geyer}
 and \citeasnoun{knight}, which states that if (i) a sequence of convex lower-semicontinous
 function $\QQ_{T}: \Bbb{R}^d \to \bar{\Bbb{R}}$
 converges in distribution in the finite-dimensional sense to  $\QQ_{\infty}: \Bbb{R}^d \to \bar{\Bbb{R}}$
over a dense subset of $\Bbb{R}^d$, (ii) $\QQ_{\infty}$ is finite over
a non-empty open set $\mathcal{Z}_0 \subset \Bbb{R}^d$, and (iii) $\QQ_{\infty}$ is
uniquely minimized at
 a random vector $\widetilde Z_{\infty}$, then any argmin of $\QQ_{T}$,
 denoted $\widetilde Z_T$, converges in distribution to $\widetilde Z_{\infty}$.

By the Convexity lemma we conclude that $\widetilde Z_{\sss T}(k)\in \arg\min_{z \in \Bbb{R}^d}
\widetilde{Q}_T(z,k) $ converges in distribution to $\widetilde
Z_{\infty}(k) =  \arg\min_{z \in \Bbb{R}^d}
\widetilde{Q}_\infty(z,k),$ where the random vector
$\widetilde Z_{\infty}(k)$ is uniquely defined  by Lemma 1 in Appendix E.

\textsc{Step 3}. By C1, $ \aT \( \beta (\tau) - \beta_e\)$ $\to$
$k^{-\xi} \gamma$ as $\tau T \to k$ and $T \to \infty$.  Thus $ \aT
( \hat \beta (\tau) - \beta(\tau)) \ind \hat Z_{\infty}(k) :=
\widetilde Z_{\infty}(k) + k^{-\xi}\gamma. $ Then $$ \hat
Z_{\infty}(k) = \widetilde Z_{\infty}(k) +k^{-\xi}\gamma =
\arg\min_{z \in \Bbb{R}^d} [ - k E[X] '(z +k^{-\xi}\gamma ) +
\sum_{t=1}^{\infty} \ell (J_t , \XX_t'(z + k^{-\xi}\gamma)) ].$$

\textsc{Step 4.} Similarly to step 2 it follows that $$\big (
\widetilde Z_T(m k), \widetilde Z_T( k) \big ) \in \text{ argmin
}_{(z_1',z_2')'  \in \mathbb{R}^{2d}} \ \QQ_T(z_1, mk)  + \QQ_T(z_2,
k)$$ weakly converges to $$\big ( \widetilde Z_\infty(mk), \widetilde
Z_\infty(k)) = \text{ argmin }_{(z_1', z_2')' \in \mathbb{R}^{2d}} \
\QQ_\infty(z_1, mk) + \QQ_\infty(z_2, k),$$
where the random vectors
$\widetilde Z_{\infty}(k)$ and $\widetilde Z_{\infty}(mk)$  are uniquely defined  by Lemma 1 in Appendix E.

Therefore it follows that
$$
\left (\widehat Z_T(k) , \frac{\mathcal{A}_T}{A_T} \right) = \left
(\widehat Z_T(k) , \frac{\bar X_T'(\widetilde Z_T(km) - \widetilde
Z_T(k) )}{\sqrt{\tau T}} \right) \to_d  \left( \widehat Z_\infty(k)
, \frac{E[X]'(\widetilde Z_\infty(km) - \widetilde Z_\infty(k)
}{\sqrt{k}} \right).
$$
By Lemma 1 in Appendix E, $E[X]'(\widetilde Z_{\infty}(mk)-
\widetilde Z_{\infty}(k)) \neq 0$ a.s., provided that $mk- k> d$. It
follows by the Extended Continuous Mapping Theorem that
\begin{eqnarray*} & & Z_T(k) = \frac{\mathcal{A}_T}{A_T} \widehat
Z_T(k)  = \frac{\sqrt{\tau T} \widehat Z_T(k)}{\bar X_T'(\widetilde
Z_T(mk) - \widetilde Z_T(k))}   \ind Z_{\infty }(k) = \frac{\sqrt{k}
\widehat Z_\infty(k)}{ E[X]'(\widetilde Z_\infty(mk) -  \widetilde
Z_\infty(k))}.
  \end{eqnarray*}
Using the relations $\widehat Z_\infty(k) = \widetilde Z_{\infty}(k)
+ k^{-\xi}\gamma$ and   $\widehat Z_\infty(mk) = \widetilde
Z_{\infty}(mk) + (mk)^{-\xi}\gamma$  and
 $E[X]'\gamma=1$ holding by C1, we can represent
 $$
 Z_{\infty }(k) = \frac{\sqrt{k} \widehat
Z_\infty(k)}{ E[X]'(\widehat Z_\infty(mk) -  \widehat Z_\infty(k)) +
(m^{-\xi} -1)k^{-\xi} }.\qed
 $$

\section{Proof of Theorem 2 and 3}

The results follows by Theorem 1 and the definition of convergence
in distribution. \qed

\section{Proof of Theorem 4 and 5}

We will prove Theorem 4. The proof of Theorem 5 follows similarly.
The main step of the proof, step 1, is specific to our problem. Let
$ G_{\sss T} (x) := Pr \{ V_{\sss T} \leq x \}$  and $G(x) := Pr \{
V_{\sss \infty} \leq x \}=\limsup_{\sss T \to
\infty} G_{\sss T}(x).$ \\
 \textsc{ Step 1.}  Letting $ V_{i, b, \sss T}:=
\mathcal{A}_{i, b, \sss T} \psi'\( \hat \beta_{i, b, \sss
T}(\tau_{\sss b}) - \beta(\tau_{\sss b})\)$, define
 \begin{eqnarray*}
 \hat G_{b, \sss T}(x) & := & B_{\sss T}^{-1} \sum_{i =1}^{B_{\sss
T}} 1\{ \hat V_{i, b, \sss T} \leq x\}  =  B_{\sss T}^{-1} \sum_{i
=1}^{B_{\sss T}} 1 \Big\{V_{i, b, \sss T}  + \mathcal{A}_{i, b, \sss
T} \psi' \( \beta(\tau_{\sss b}) - \hat \beta(\tau_{\sss b})\)
\leq x \Big \}, \\
\dot G_{b, \sss T}(x; \Delta) & := & B_{\sss T}^{-1} \sum_{i
=1}^{B_{\sss T}} 1\{ V_{i, b, \sss T} + \(\mathcal{A}_{i, b, \sss
T}/A_b\)\times \Delta \leq x \},
\end{eqnarray*}
where $A_b=1/Q_U(1/b)$ is the canonical normalizing constant.
 Then
\bs
 & 1  [ V_{i, b, \sss T} \leq x - \mathcal{A}_{i, b, \sss T}w_T/ A_b
 ]  \leq 1  [\hat V_{i, b, \sss T} \leq x
 ] \leq  1  [V_{i, b, \sss T} \leq x + \mathcal{A}_{i, b, \sss T}
 w_T/A_b
 ]
\end{split}\end{align}
for all $i =1,..., B_{\sss T}$, where $ w_T = | A_b
\psi'(\beta(\tau_{\sss b}) - \hat \beta(\tau_{\sss b})) |.$

The principal claim is  that, under conditions of Theorem 3, $ w_T =
o_p(1).$  The claim follows by noting that for $k_{\sss
T}=\tau_{\sss T} T \to k>0$  as $b/T \to 0$ and $T \to \infty$,
 \bsnumber \label{pt3order1} & A_b \times
\(\beta(\tau_{\sss b}) - \hat \beta(\tau_{\sss b})\)  \sim
 \frac{ 2^{-\xi}-1}{ Q_U(2 k_{\sss T}/b) - Q_U(k_{\sss T}/b)}  \times O_p \(
\frac{Q_U(2 k_{\sss T}/b) - Q_U(k_{\sss T}/b)}{\sqrt{ \tau_b \cdot
T}}
\) \\
& = O_p\( \frac1{\sqrt{ \tau_b \cdot T}}   \)  = O_p\(
\sqrt{\frac{b}{k_{\sss T} T}}\) =  O_p\(
\sqrt{\frac{b}{k T}}\)  = o_p(1)
\end{split}\end{align}
The first relation in (\ref{pt3order1}) follows from two facts:
First, by definition $A_b := 1/Q_U(1/b) $ and by the regular
variation of $Q_U$ at $0$ with exponent $-\xi$, for any $l \searrow
0$, $Q_U(l) (2^{-\xi} - 1) \sim Q_U(2l) - Q_U(l).$ Second, since
$\tau_b = k_{\sss T}/b$ and since $\tau_b \times T = (k_{\sss
T}/b)\times T \sim (k/b)\times T \to \infty$ at a polynomial speed
in $T$ by $T/b \to \infty$ at a polynomial speed in $T$ by
assumption, $\hat \beta(\tau_{\sss b})$ is the intermediate order
regression quantile computed using the \textit{full sample} of size
$T$, so that by Theorem 3 in \citeasnoun{victor:annals}
 \bsnumber
\beta(\tau_{\sss b}) - \hat \beta(\tau_{\sss b}) = O_p\(
\frac{Q_U(2k_{\sss T}/b ) - Q_U( k_{\sss T}/b)}{\sqrt{\tau_b \cdot
T} } \).
 \end{split}\end{align}

Given that $w_T =o_p(1)$, for some sequence of constants
$\Delta{\sss T} \searrow 0$ as $T \to \infty$ the following event
occurs wp $\rightarrow 1$ : \bs M_{\sss T} =\left\{
\begin{array}{lll} 1  [ V_{i, b, \sss T} < x - \mathcal{A}_{i, b,
\sss T} \Delta{\sss T}/ A_b
 ]  & \leq 1  [ V_{i, b, \sss T} < x - \mathcal{A}_{i, b, \sss T}w_T/ A_b
 ] \\ & \leq 1  [\hat V_{i, b, \sss T} < x
 ] \\ & \leq  1  [V_{i, b, \sss T} < x + \mathcal{A}_{i, b, \sss T}
 w_T/A_b
 ] \\ & \leq 1  [ V_{i, b, \sss T} < x + \mathcal{A}_{i, b, \sss T}\Delta{\sss T}/ A_b
 ], \\ & \text{ for all $i=1,..., B_{\sss T}$ }. \end{array} \right \} .
\end{split}\end{align}
Event $M_{\sss T}$ implies
 \bsnumber\label{Gb}
 \dot G_{b, \sss T}(x;
\Delta{\sss T}) \leq \hat G_{b, \sss T}(x)  \leq \dot G_{b, \sss T}(x;
-\Delta{\sss T}).
 \end{split}\end{align}

\textsc{ Step 2.}  In this part we show that at the continuity
points of $G(x)$, $\dot G_{b, \sss T}(x; \pm\Delta{\sss T}) \inp
G(x)$. First, by non-replacement sampling  \be E [\dot G_{b, \sss
T}(x; \Delta{\sss T})]= P [ V_{b} - \mathcal{A}_{b} \Delta{\sss T}/ A_b
\leq x ].\ee Second, at the continuity points of $G(x)$
 \bsnumber\label{contin}
\lim_{\sss T \to \infty} E [\dot G_{b, \sss T}(x; \Delta{\sss T})] =
\lim_{ b \to \infty } P [ V_{ b} - \mathcal{A}_{b} \Delta{\sss T}/ A_b
\leq x ]  =  P [ \psi'Z_{\infty}(k) \leq x ]  = G(x).
\end{split}\end{align}
 The statement (\ref{contin}) follows because
$
 V_{
b} - \frac{\mathcal{A}_{b} \Delta{\sss T}}{A_b} = V_b + o_p(1) \ind
\psi'Z_{\infty}(k) , $ since by Theorem 1 $V_b \ind
\psi'Z_{\infty}(k)$ and by the proof of Theorem 1 and by $\Delta_{\sss T} \searrow 0$
 $$\frac{\mathcal{A}_{b} \Delta{\sss T}}{A_b} = O_p(1) \cdot \Delta{\sss T} = O_p(1) \cdot o(1) = o_p(1).$$

Third, because  $\dot G_{b, \sss T}(x, \Delta{\sss T})$ is a
U-statistic of degree $b$,  by the LLN for U-statistics in
\citeasnoun{romano:book}, $ Var(\dot G_{b, \sss T}(x, \Delta{\sss T}))
= o(1) .
 $
This shows that $\dot G_{b, \sss T}(x; \Delta{\sss T}) \inp G(x)$. By
the same argument $\dot G_{b, \sss T}(x; -\Delta{\sss T}) \inp
G(x)$.

\textsc{Step 3.} Finally,  since  event $M_{\sss T}$ occurs wp $\to
1$ and so does (\ref{Gb}), by  Step 2 it follows that $\hat G_{b,
\sss T}(x) \inp G(x)$ for each $x \in \Bbb{R}$. Finally,  convergence of distribution functions at continuity points,
implies convergence of quantile functions at continuity points. Therefore, by
the Extended Continuous Mapping Theorem,
$ \hat c_{\alpha}=\hat G^{-1}_{b, \sss T}(\alpha) \inp
c_{\alpha}=G^{-1}(\alpha), $ provided $G^{-1}(\alpha)$ is a
continuity point of $G(x)$.    \qed

\section{Proof of Theorems 6 and 7}

We will prove Theorem 7; the proof of Theorem 6 follows similarly.

We prove the theorem  by showing that the law of the limit variables
is continuous in the underlying parameters, which implies the
validity of the proposed procedure.  This proof structure is similar
to the one used in the parametric bootstrap proofs, with the
complication that the limit distributions here are  non-standard.
The demonstration of continuity poses some difficulties, which we
deal with by invoking epi-convergence arguments and exploiting the
properties of the Poisson process (\ref{eq: limit N}). We also carry
out the proof for the case with $\xi<0$; the proof for the case with
$\xi>0$ is identical apart from a change in sign in the definition
of the points of the Poisson process, as indicated in the proof of
Theorem 1.

Let us first list the basic objects with which we will work:

\textbf{1.} The parameters are $\xi \in (-\infty,0)$, $\gamma \in
\Bbb{R}^d$, and $F_X \in \mathcal{F}_X$, a distribution function on
$\Bbb{R}^d$ with the compact support $\mathbf{X}$. We have the set
of estimates such that: \be \sup_{x\in \mathbf{X}}|F_X(x) - \hat
F_X(x)| \to_p 0, \ \hat \xi \to_p \xi, \ \hat \gamma \to_p \gamma
\text{ as } T \to \infty, \ee where $\widehat F_X \in
\mathcal{F}_X$.  The set $\mathcal{F}_X$ is the set of non-lattice
distributions defined in Appendix \ref{unique}.
 The underlying probability space $(\Omega, \mathcal{F}, P)$ is
the original probability space induced by the data.

\textbf{2.} $\NN$ is a Poisson random measure (PRM), with mean
intensity measure $m_\NN$, and points representable as: $
(\Gamma^{-\xi}_j \cdot \mathcal{X}_j '\gamma, \ \ \ \mathcal{X}_j),
\ \ j =1,2,3, ... . $ $\NN$ is a random element of a complete and
separable metric space of point measures $(M_p(E), \rho_v)$ with
metric $\rho_v$ generated by the topology of vague convergence. The
underlying probability space $(\Omega', \mathcal{F}', P')$ is the
one induced by Monte-Carlo draws of points of $\NN$. This law of
$\NN$ in $(M_p(E), \rho_p)$ will be denoted as $\mathcal{L}(\NN
|\xi, \gamma, F_X)$. The law depends only on the parameters $(\xi,
\gamma, F_X)$ of the intensity measure $m_{\mathbf{N}}$.

\textbf{3.} The random objective function (ROF)  takes the form
\bsnumber z \mapsto \widehat Q_{\infty}(z;k) & =  - k E[X] '
(z+k^{-\xi}\gamma) + \int_E \{x'(z + k^{-\xi}\gamma) - u^{-\xi}\cdot x'\gamma \}_+ d \NN(u, x) \\
&  = - k E[X]' (z+k^{-\xi}\gamma) + \ \sum_{t=1}^{\infty} \{
\XX_t'(z +k^{-\xi}\gamma) -\Gamma_t^{-\xi}\cdot \XX_t'\gamma \}_+,
\end{split}\end{align} and is a random element of the metric space of
proper lower-semi-continuous functions $(LC(\Bbb{R}^d), \rho_{e})$,
equipped with the metric $\rho_e$ induced by the topology of
\textit{epi-convergence}. \citeasnoun{geyer} and \citeasnoun{knight}
provide a detailed introduction to epi-convergence, with connections
to convexity and stochastic equi-semicontinuity. Moreover, this
function is convex in $z$, which is a very important property to
what follows.
The law of $z \mapsto \widehat Q_{\infty}(z;k)$ in $(LC(\Bbb{R}^d),
\rho_{e})$ will be denoted as $\mathcal{L}(\widehat
Q_{\infty}(\cdot;k) | \xi, \gamma, F_X)$. This law depends only on
the parameters $(\xi, \gamma, F_X)$.

\textbf{4.} The extremum statistic $ \widehat Z_{\infty}( k) = \arg
\min_{z \in \Bbb{R}^d} \widehat Q_{\infty}(z;k)$ is a random element
in the metric space $\Bbb{R}^d$, equipped with the usual Euclidian
metric.
The law of $\widehat Z_{\infty}(k)$ in $\Bbb{R}^d$ will be denoted
as $\mathcal{L}(\widehat Z_{\infty}(k)| \xi, \gamma, F_X)$. This law
depends only on the parameters $(\xi, \gamma, F_X)$.

Next we collect together several weak convergence properties of the
key random elements, which are most pertinent to establishing the
final result.

\textbf{A.} A sequence of PRM $(\NN^m, m =1,2, ...)$ in $(M_p(E),
\rho_p)$ defined by the sequence of  intensity measures $m_{\NN^m}$
with parameters $(\xi^m, \gamma^m, F_X^m)$ converges weakly to a PRM
$\NN$ with intensity measure $m_\NN$ with parameters $(\xi, \gamma,
F_X)$ if the law of the former converges to the law of the latter
with respect to the Bounded-Lipschitz metric $\rho_w$ (or any other
metric that metrizes weak convergence): \be\label{Nconvergence}
\lim_{m \to \infty} \rho_{w} ( \mathcal{L}(\NN^m|\xi^m, \gamma^m,
F_X^m), \mathcal{L}(\NN|\xi, \gamma, F_X) ) = 0.
 \ee
The weak convergence of PRMs is equivalent to pointwise convergence
of their Laplace functionals: \be\label{laplaceconverge} \lim_{m \to
\infty} \varphi(f; \NN^m) = \varphi(f; \NN), \ \ \forall f \in
C^+_K(E),\ee where $C^+_K(E)$ is the set of continuous positive
functions $f$ defined on the domain $E$ and vanishing outside a
compact subset of $E$. The Laplace functional is defined as and
equal to:
 \be
\varphi(f; \NN) &:= E \[e^{\int_E f(u, x) \ d \NN(u,x)} \]  = e^{\(
- \int_E
\[1- e^{- f(u, x)}\] d m_\NN(u, x)\)}. \ee

\textbf{B.} A sequence of ROFs $\{\widehat Q^m_{\infty}(\cdot;k),
m=1,2,3,...\}$ defined by the sequence of parameters $\{(\xi^m,
\gamma^m, F_X^m), m =1,2,3,...\}$  converges weakly to the ROF
$\widehat Q_{\infty}(\cdot;k)$ defined by parameters $(\xi, \gamma,
F_X)$ in the metric space $(LC(\Bbb{R}^d), \rho_{e})$, if the law of
the former converges to the law of the latter with respect to the
Bounded Lipschitz metric  $\rho_w $(or any other metric that
metrizes weak convergence): \be\label{epiconvergence} \lim_{m \to
\infty} \rho_w ( \mathcal{L}( \widehat Q^m_{\infty}(\cdot;k) |\xi^m,
\gamma^m, F_X^m), \mathcal{L}(\widehat Q_{\infty}(\cdot;k)| \xi,
\gamma, F_X) ) = 0.
 \ee
Moreover, since the objective functions are convex in $z$, the above
weak convergence is equivalent to the finite-dimensional weak
convergence:
 \be\label{fidiconvex}
(\widehat Q_{\infty}^m(z_j;k), j=1,...,J) \to_d ( \widehat
Q_{\infty}(z_j;k), j =1,...J) \ \ \ \text{ in } \Bbb{R}^{J}
 \ee
for any finite collection of points $(z_j, j =1,...,J)$.  The result,
that finite-dimensional convergence in distribution of convex functions
implies epi-convergence in distribution, is due to \citeasnoun{geyer} and
\citeasnoun{knight}. Thus, in order to check (\ref{epiconvergence}) we only
need to check (\ref{fidiconvex}).

\textbf{C.} In turn, the weak convergence of objective functions
$\{\widehat Q^m_{\infty}(\cdot;k), m=1,2,3,...\}$ to $\{\widehat
Q_{\infty}(z\cdot;k)$ in $(LC(\Bbb{R}^d), \rho_{e})$ implies that as
$m \to \infty$ the weak convergence of the corresponding argmins: $
\widehat Z_{\infty}^m(k) \to_d \widehat Z_{\infty}(k)$  in
$\Bbb{R}^{d},$
 that is,
 \be\label{argminconv}
\lim_{m \to \infty}  \rho_w ( \mathcal{L}( \widehat
Z_{\infty}(k)|\xi^m, \gamma^m, F_X^m), \mathcal{L}(\widehat
Z_{\infty}(k)|\xi, \gamma, F_X) ) = 0. \ee

The proof is now done in two steps:

\textbf{I.} We would like to show that the law $\mathcal{L}(
\widehat Z_{\infty}(k)| \xi', \gamma', F_X')$ is continuous at
$(\xi', \gamma', F_X') = (\xi, \gamma, F_X)$ for each $(\xi, \gamma,
F_X)$ in the parameter space, that is, for any sequence $ (\xi^m,
\gamma^m, F_X^m , m=1,2,...)$ such that \be\label{parameterconv} |
\xi^m - \xi| \to 0, |\gamma^m - \gamma| \to 0, \sup_{x \in
\mathbf{X}} | F^m_X(x) - F_X(x) | \to 0\ee with $F_X^m \in
\mathcal{F}_X$, we have \be\label{keycontinuity}
 \rho_w ( \mathcal{L}( \widehat Z_{\infty}(k)| \xi^m,
\gamma^m, F_X^m), \mathcal{L}(\widehat Z_{\infty}(k)| \xi, \gamma,
F_X)) \to 0. \ee

\textbf{II.} Given this continuity property, as $ | \hat \xi - \xi|
\to_p 0, |\hat \gamma - \gamma| \to_p 0, \sup_{x \in \mathbf{X}} |
\hat F_X(x) - F_X(x) | \to_p 0,$ we have by the Continuous Mapping
Theorem \be
 \rho_w ( \mathcal{L}( \widehat Z_{\infty}(k)| \hat
\xi, \hat \gamma, \hat F_X), \mathcal{L}(\widehat Z_{\infty}(k)|
\xi, \gamma, F_X) ) \to_p 0. \ee That is, the law  $\mathcal{L}(
\widehat Z_{\infty}(k)| \hat \xi, \hat \gamma, \hat F_X)$ generated
by the Monte Carlo procedure consistently estimates the limit law
$\mathcal{L}(\widehat Z_{\infty}(k)| \xi, \gamma, F_X)$, which is
what we needed to prove, since this result implies that  the
convergence of respective distribution functions at the continuity
points of
 the limit distribution. Convergence of the distribution functions at the continuity points of
 the limit distribution implies convergence of the respective
 quantiles to the quantiles of the limit distribution provided the
 latter are positioned at the continuity points of the limit distribution function.

Thus it only remains to show the key continuity step I.  We have
that {\small
 \be \nonumber
(\ref{parameterconv})\overset{(1)}\implies (\ref{laplaceconverge})
\overset{(2)}\implies (\ref{Nconvergence})  \overset{(3)} \implies
(\ref{fidiconvex}) \overset{(4)} \implies (\ref{epiconvergence})
\overset{(5)}\implies (\ref{argminconv}) \Leftrightarrow
(\ref{keycontinuity}) \ee } where (1) follows by direct
calculations: for $g(u,x) = 1 - e^{- f(u,x)}$ and any $f \in C_K(E)$
\bs |   \varphi(f; \NN^m) -\varphi(f; \NN) |  \leq \varphi(f; \NN)
\left | \exp\left \{ \int_E g(u,x) [dm_{\NN} - dm_{\NN^m}] \right\}
- 1 \right | \to 0,
\end{split}\end{align}
as $\int_E \( g(u,x) \[dm_{\NN} - dm_{\NN^m}\]\) \to 0$, which
follows from the definition of the measure $m_{\NN}$ stated earlier;
(2) follows by the preceding discussion in Step A;  (3) follows by the continuity of
the mapping $$\NN \mapsto \int_E ( x'(z +k^{-\xi}) -u^{-\xi}\cdot
x'\gamma)_+ d \NN(u, x) $$ from $(M_p(E), \rho_v)$ to $\Bbb{R}$,
as noted in the proof of Theorem 1; and (4)
and (5) follow by the preceding discussion in Step C. \qed

\section{Uniqueness and Continuity}\label{unique}

Define $k:= \lim_{T \to \infty} \tau T$ and fix an $m$ such that $k(m - 1)> d$ where $d = \text{dim}(X)$.

 Let  $\{\XX_t, t \geq 1\}$ be an i.i.d. sequence from a distribution function $F_X$ such that $E
[\XX \XX']$ is positive definite. Define $G_j := (kE[X] - \sum_{t
\leq  j } \XX_t)' \{[\XX_{j +1} ... \XX_{j +d}]'\}^{-1}$ if the
matrix $[\XX_{j +1} ... \XX_{ j+d}]'$ is invertible, and $G:=
(\infty,...,\infty)$ otherwise. Denote by $\mathcal{F}_X(k)$ the
class of distributions $F_X$ for which   $ P_{F_X}\{ G_j \in
\partial
(0,1)^d \}=0$ for all integer $j \geq 0$.  \\

\textbf{Definition} (Non-Lattice Condition Given $k$ and $m$).
 $F_X \in \mathcal{F}_X(k')$
for both $k'=k$ and $k'=mk$.

Denote the class of all non-lattice distributions as  $\mathcal{F}_X
= \mathcal{F}_X(k) \cap \mathcal{F}_X(mk)$.

\begin{lemma} If $F_X \in \mathcal{F}_X$,  then
$\widehat Z_{\infty}(k)$ and $Z_{\infty}(k)$ are uniquely defined random vectors. Moreover,
for any $\psi \neq 0$,  $\psi'\widehat Z_{\infty}(k)$ and  $\psi'Z_{\infty}(k)$ have continuous distribution functions.
\end{lemma}

\begin{remark}  The non-lattice condition is an analog of Koenker and
Bassett's (1978) condition for uniqueness of quantile regression in
finite samples. This condition trivially holds if the nonconstant
covariates $\mathcal{X}_{-1t}$ are absolutely continuous. Uniqueness
therefore holds generically in the sense that for a fixed $k$ adding
arbitrarily small absolutely continuous perturbations to
$\{\mathcal{X}_{-1t}\}$ ensures it.
 \end{remark}

\textsc{Proof:}  \textsc{Step 1. }We have from Theorem 1 that
$\widehat Z_{\infty}(k)=\widetilde Z_{\infty}(k)+ c$ for some
constant $c$, where $\widetilde Z_{\infty}$ is defined in Step 2 of
the proof of Theorem 1. \citeasnoun{victor:annals} shows that a
sufficient condition for tightness of possibly set-valued
$\widetilde Z_{\infty}(k)$ is $E [\XX\XX']>0$.  Taking tightness as
given, conditions for uniqueness and continuity of $\widetilde
Z_{\infty}(k)$ can be established. Define $\mathcal{H}$ as the set
of all $d$-element subsets of $\mathbb{N}=\{1,2,3,...\}$. For $h \in
\mathcal{H}$, let $\XX(h)$ and $J(h)$ be the matrix with rows
$\{\XX_t, t \in h\}$, and vector with elements $\{J_t, t \in h\}$,
respectively, where $J_t$ are defined in the proof of Theorem 1. Let
$\mathcal{H}^* =\{ h \in \mathcal{H}: |\XX(h)| \neq 0 \}$. Nota that
$\mathcal{H}^*$ is non-empty a.s. by $E[\XX\XX']$ positive definite
and is countable. By the same argument as in the proof of Theorem
3.1. of \citeasnoun{koenker:1978} at least one element of
$\widetilde{Z}_{\infty}(k)$ takes the form $z_h= \XX(h)^{-1}J(h)$
for some $h \in \mathcal{H}^*$, and must satisfy a sub-gradient
condition: $$\mathbf{\zeta}_{k}(z_h) := ( k E[X] -
\sum_{t=1}^{\infty}
 1(J_t < \XX_t' z_h) \XX_t) '\XX(h)^{-1} \in [0,1]^d,$$
and the argmin is unique if and only if $\mathbf{\zeta}_{k}(z_h) \in
\mathcal{D} = (0,1)^d.$  By the same argument as in the proof of
Theorem 3.4 in  \citeasnoun{koenker:1978},  $z_h$ must obey
$$
k -d \leq \sum_{t=1}^{\infty}
 1(J_t < \XX_t' z_h) \leq  k.
$$

Then, uniqueness holds for a fixed $k>0$ if $P \(  \exists h
\in\mathcal{H}^* : \mathbf{\zeta}_{k}(z_h) \in
\partial \mathcal{D} \) =0.$   To show this is the case,   define  $\mathcal{M}(j)$ as the set of
all $j$-element subsets of $\mathbb{N}$, and define for  $\mu \in
\mathcal{M}(j)$, $ G(\mu,h) := (kE[X] - \sum_{t \in \mu} \XX_t)'
\XX(h)^{-1}$ if $\XX(h)$ is invertible, and  $G(\mu,h) :=
(\infty,...,\infty)$ otherwise. Now note that if $P_{F_X}\{ G(\mu,h)
\in
\partial \mathcal{D}\}=0$ for any $h \in \mathcal{H}$ and $\mu \in
\mathcal{M}(j)$ such that $h \cap \mu = \emptyset$ and any integer
$j \geq 0$,  then
 \bs
 & \displaystyle P \(  \exists h \in\mathcal{H}^* : \mathbf{\zeta}_{k}(z_h) \in
\partial \mathcal{D} \)  \\
&\displaystyle \leq P \{ G(\mu,h) \in \partial \mathcal{D}, \exists
h \in \mathcal{H}, \exists \mu \in
\mathcal{M}(j), \exists j \geq 0: h \cap \mu = \emptyset,  k-d \leq j \leq k\}  \\
& \displaystyle\leq  \sum_{ (k-d) \vee 0 \leq j \leq k } \ \sum_{h
\in \mathcal{H}} \ \sum_{ \mu \in \mathcal{M}(j): h \cap \mu =
\emptyset } P_{F_X}\{ G(\mu,h) \in
\partial \mathcal{D}\} = 0,
\end{split}\end{align}
since the summation is taken over the countable set.  Finally, by
the i.i.d. assumption and $h \cap \mu = \emptyset$, $P_{F_X}\{
G(\mu,h) \in
\partial \mathcal{D}\}= P_{F_X}\{ G_j \in
\partial \mathcal{D} \}$, where $G_j$ is defined above. Therefore, $P_{F_X}\{ G_j  \in
\partial \mathcal{D} \}=0$ for all integer $j \geq 0$ is the condition that suffices for
uniqueness.

\textsc{Step 2.} Next want to show that the distribution function of
$\psi'\widetilde Z_{\infty}(k)$ has no point masses, that is
$P\{\psi'\widetilde Z_{\infty}(k) =x\} = 0$ for each $x \in \Bbb{R}$
and each vector $\psi \neq 0$, which is the equivalent to showing
continuity of $x \mapsto P\{\psi'\widetilde Z_{\infty}(k) \leq x\}$
at each $x$. Indeed, from above we have that
$\{\widetilde{Z}_{\infty}(k)= \XX(h)^{-1}J(h)\}$ for some $h \in
\mathcal{H}^*$ a.s.,  and $
 P \{\psi'\XX(h)^{-1}J(h)= x|  \{\XX_t, t\geq 1 \}\} = 0
$ for each $h \in \mathcal{H}^*$ a.s., since $J(h)$ is absolute
continuous conditional on  $\{\XX_t, t\geq 1 \}$ and
$\psi'\XX(h)^{-1} \neq 0$.   Therefore,
$$
P\{\widetilde Z_{\infty}(k) =z\} \leq E \left [\sum_{h \in
\mathcal{H}^*} P \{\psi'\XX(h)^{-1}J(h)= z| \{\XX_t, t\geq 1 \}\}
\right ] = 0,
$$
 by countability of $\mathcal{H}$ and the law of iterated expectations.

\textsc{Step 3.} Next we want to show that the distribution function
of $\psi'Z_{\infty}(k)$ has no point masses, that is
$P\{\psi'Z_{\infty}(k) =x\} = 0$ for each $x \in \Bbb{R}$ and each
vector $\psi \neq 0$.
 We have that $$Z_{\infty}(k) =
\sqrt{k} (\widetilde Z_\infty(k) + c)/(E[X]'(\widetilde Z_\infty(mk)
-  \widetilde Z_\infty(k)))$$ for some $c\neq 0$. From Steps 1 and 2
we  have that solutions $\widetilde{Z}_{\infty}(mk)$ and
$\widetilde{Z}_{\infty}(k)$ are a.s. unique and a.s. take the form
$\widetilde{Z}_{\infty}(k) = z_{h_1}= \XX(h_1)^{-1} J(h_1)$ and
$\widetilde{Z}_{\infty}(mk) = z_{h_2}= \XX(h_2)^{-1} J(h_2)$ for
some $(h_1,h_2) \in \mathcal{H}^*\times \mathcal{H}^*$. Furthermore,
$m k - k> d$ implies $h_1\neq h_2$ and hence a.s. $z_{h_1} \neq
z_{h_2}$. Indeed, to see this  by Step 2 we must have the inequality
$$
k -d \leq \sum_{t=1}^{\infty}
 1(J_t < \XX_t' z_{h_1}) \leq  k \ \text{ and } \ mk -d \leq \sum_{t=1}^{\infty}
 1(J_t < \XX_t' z_{h_2}) \leq  m k,$$
so that $h_1 = h_2$ implies  $m k - k \leq d$.
Furthermore, arguing similarly to \cite{bassett:1982}'s Theorem 2.2., we observe that for $h_1$ and $h_2$ defined above
\begin{eqnarray*}
-kE[X]'z_{h_1} &+& \int_E \ell(u, x'z_{h_1}) d \NN(u,x) - (mk- k) E[X]'z_{h_2} \\
 & < & -kE[X]'z_{h_2} + \int_E \ell(u, x'z_{h_2}) d \NN(u,x)  - (mk- k) E[X]'z_{h_2} \\
 & = &  -mk E[X]'z_{h_2} + \int_E \ell(u, x'z_{h_2}) d \NN(u,x) \\
 & < &   -mk E[X]'z_{h_1} + \int_E \ell(u, x'z_{h_1}) d \NN(u,x) \\
& <&   -k E[X]'z_{h_1} + \int_E \ell(u, x'z_{h_1}) d \NN(u,x) -
(mk-k) E[X]'z_{h_1}.
\end{eqnarray*}
Solving this inequality we obtain  $ (mk-k) E[X]' (z_{h_2} -
z_{h_1}) >0$.   We conclude therefore that a.s. $E[X]'(
\widetilde{Z}_{\infty}(mk) - \widetilde{Z}_{\infty}(k))=
E[X]'\XX(h_2)^{-1}J(h_2)- E[X]'\XX(h_1)^{-1}J(h_1) > 0$. Moreover,
conditional on $\{\mathcal{X}_t\, t \geq 1\}$ we can show by a
perturbation argument that $h_1 \neq h_2$ must be such that
$E[X]'(\widetilde{Z}_{\infty}(mk) - \widetilde{Z}_{\infty}(k))= c_1'
J(h_2) + c_2'J(h_1 \setminus h_2)$ for some
 constant  $c_2\neq 0$.  Let us denote by $\mathcal{G} $ the set of all  pairs $h_1 \neq h_2$ in $\mathcal{H}^*\times \mathcal{H}^*$ that obey these two conditions.

From step 1 and from $E[X]'( \widetilde{Z}_{\infty}(mk) -
\widetilde{Z}_{\infty}(k)) > 0$ a.s., it follows that
$Z_{\infty}(k)$ is a proper random variable. Furthermore, for any $x
\in \Bbb{R}$ a.s. for any $(h_1,h_2) \in \mathcal{G}$ and $S(h_1,
h_2)= E[X]'\XX(h_2)^{-1}J(h_2)- E[X]'\XX(h_1)^{-1}J(h_1)$, $
 P \{\psi'(\XX(h_1)^{-1}J(h_1)+c)/S(h_1, h_2)  = x \cap (h_1 \times h_2) \in \mathcal{G}| \{\XX_t, t \geq 1\} \} = 0.
$ The claim follows because, for any $(h_1,h_2) \in \mathcal{G}$,
$\psi'(\XX(h_1)^{-1}J(h_1)+c)/S(h_1, h_2)$ is  absolutely continuous
conditional on $\{\XX_t, t \geq 1\}$  by $\psi'\XX(h_1)^{-1}J(h_1)$
and $S(h_1, h_2)$ being jointly absolutely continuous  conditional
on $\{\XX_t, t \geq 1\}$ and by the non-singularity of
transformation $(w,v) \mapsto [w+c]/v$ over region $v > 0$.
Therefore, for any $x \in \Bbb{R}$,  $P\{ \psi'Z_{\infty}(k) = x \}$
is bounded above by
$$
E \left [\sum_{h_1 \neq h_2 \in \mathcal{H}^* \times \mathcal{H}^*} P
\{\psi'\XX(h_1)^{-1}J(h_1)/S(h_1, h_2) = z \cap (h_1 \times h_2) \in
\mathcal{G} |  \{\XX_t, t \geq 1\}\}  \right ] = 0,
$$
by countability of $\mathcal{H}$ and the law of iterated expectations.

\footnotesize

\linespread{.9}

\bibliography{c:/aaa/biblio/my,my,evt,rb}
\bibliographystyle{econometrica}

\newpage

\begin{figure}[htbp!] \label{sim1}
 \begin{center}
\noindent
 \centering\epsfig{figure=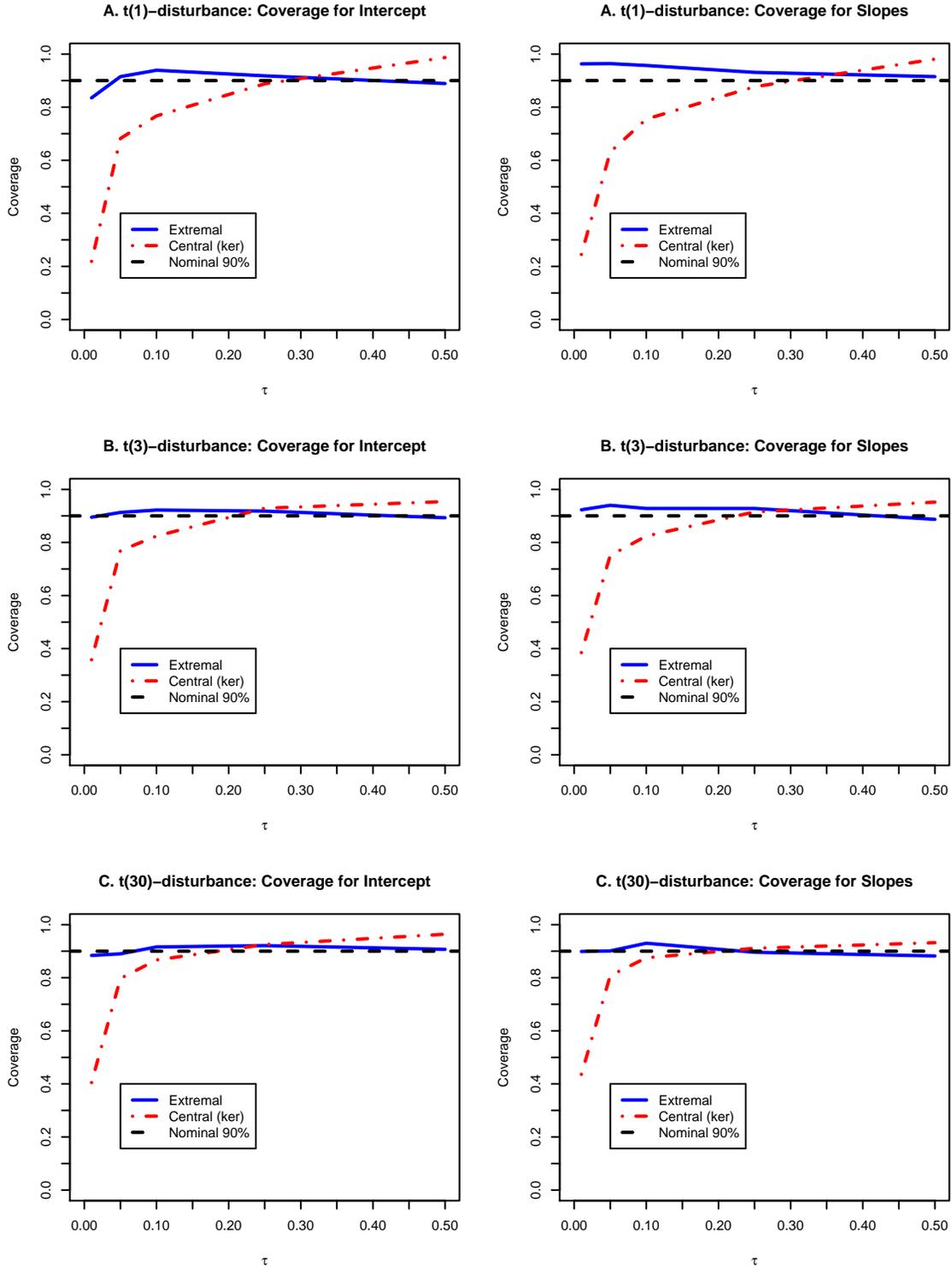, width=6in,height=8in}
\caption{Coverage of  extremal confidence intervals and normal
confidence intervals when Disturbances are $t (\nu),  \nu \in
\{1,3,30\}$. Based on 1,000 repetitions.}
 \end{center}
\end{figure}

\begin{figure}[htbp!] \label{sim1}
 \begin{center}
\noindent
 \centering\epsfig{figure=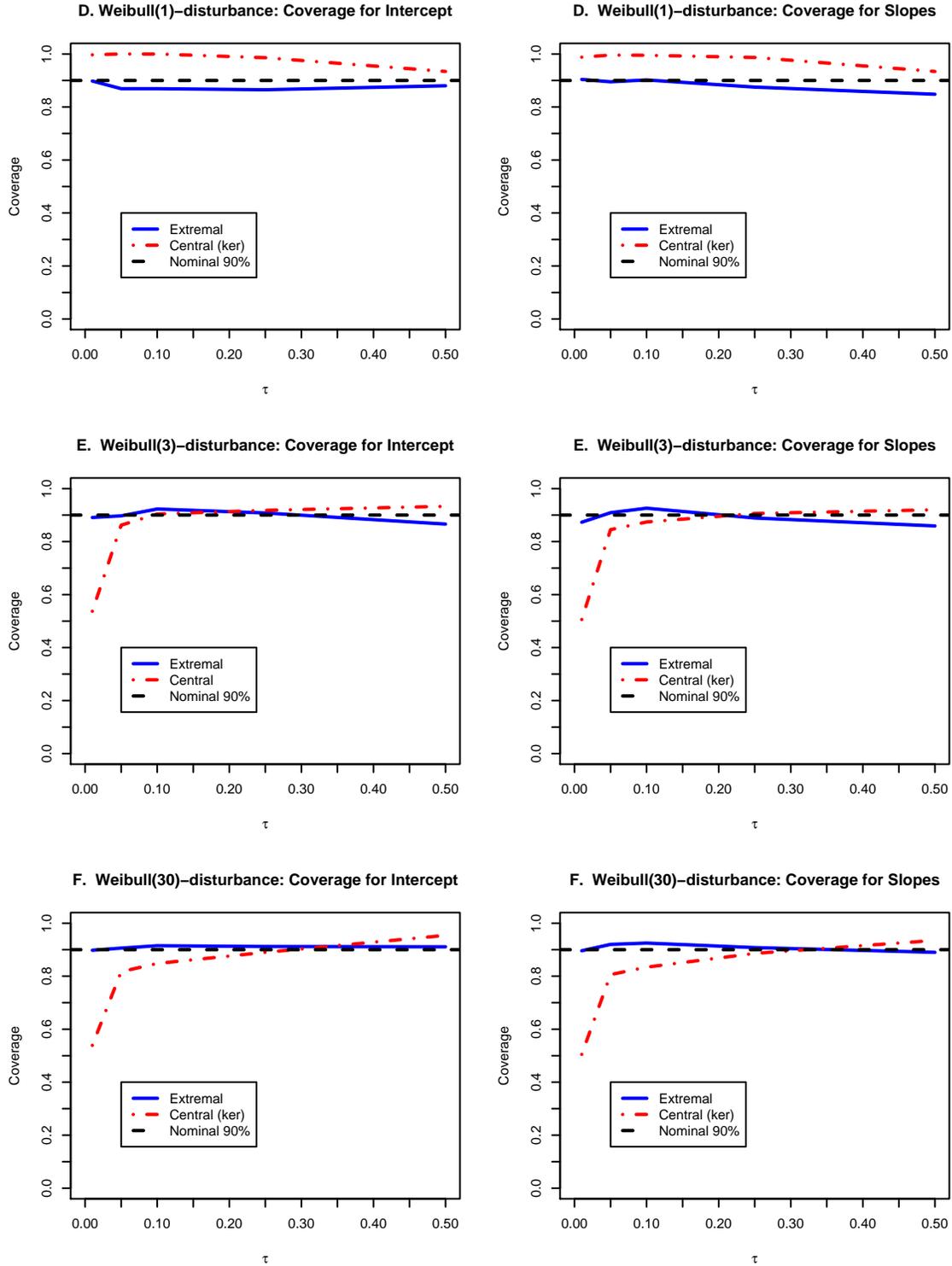, width=6in,height=8in}
\caption{Coverage of  extremal confidence intervals and normal
confidence intervals when disturbances are Weibull $(\alpha), \alpha
\in \{1,3,30\}$. Based on 1,000 repetitions.}
 \end{center}
\end{figure}


\begin{figure}[htbp!] \label{result1}
 \begin{center}
\noindent
 \centering\epsfig{figure=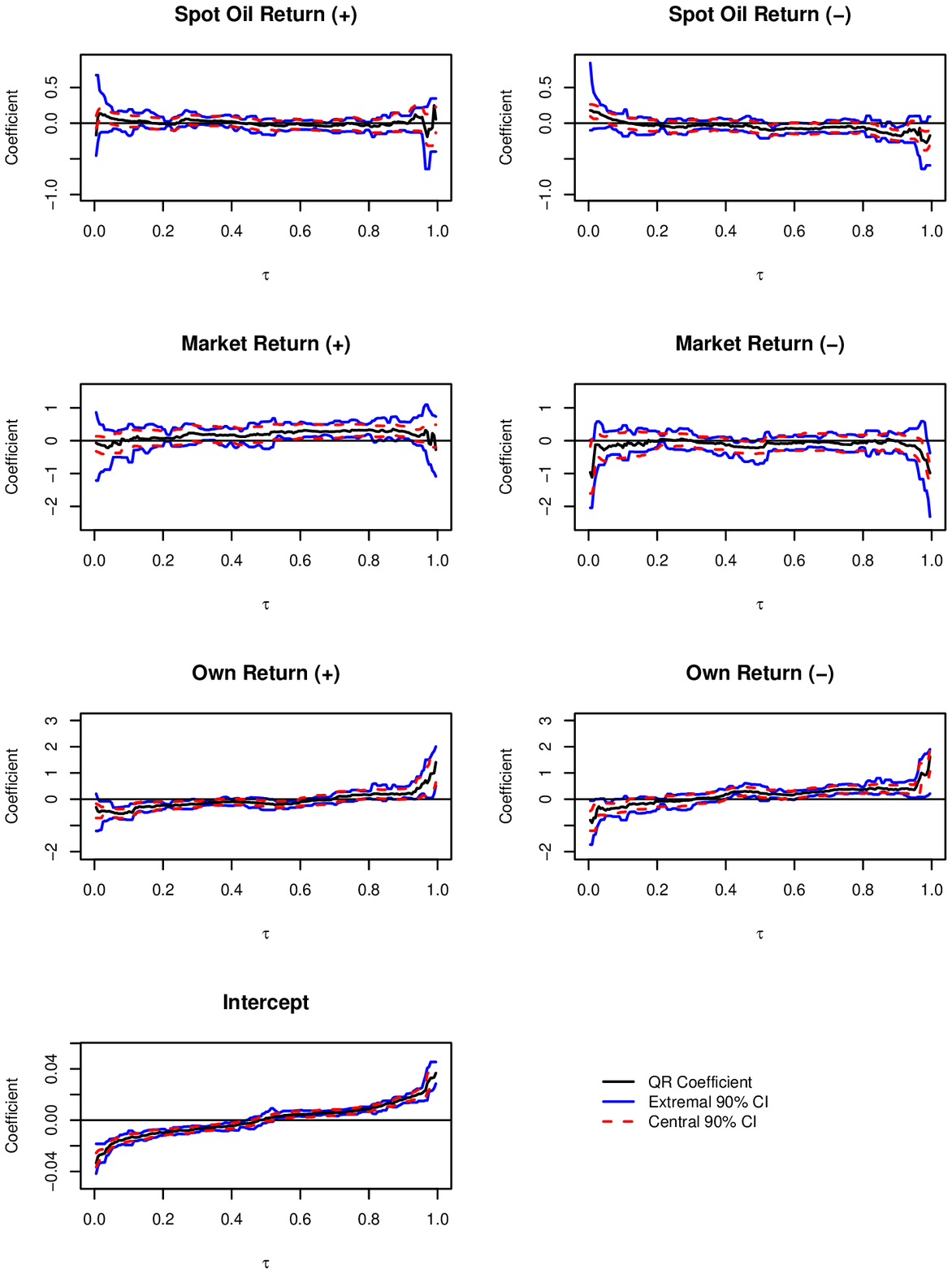, width=6in,height=8in}
\caption{QR coefficient estimates and $90\%$ pointwise confidence
intervals. The solid lines depict extremal confidence intervals. The
dashed lines depict normal confidence intervals. }
 \end{center}
\end{figure}

\begin{figure}[htbp!] \label{result2}
 \begin{center}
\noindent
 \centering\epsfig{figure=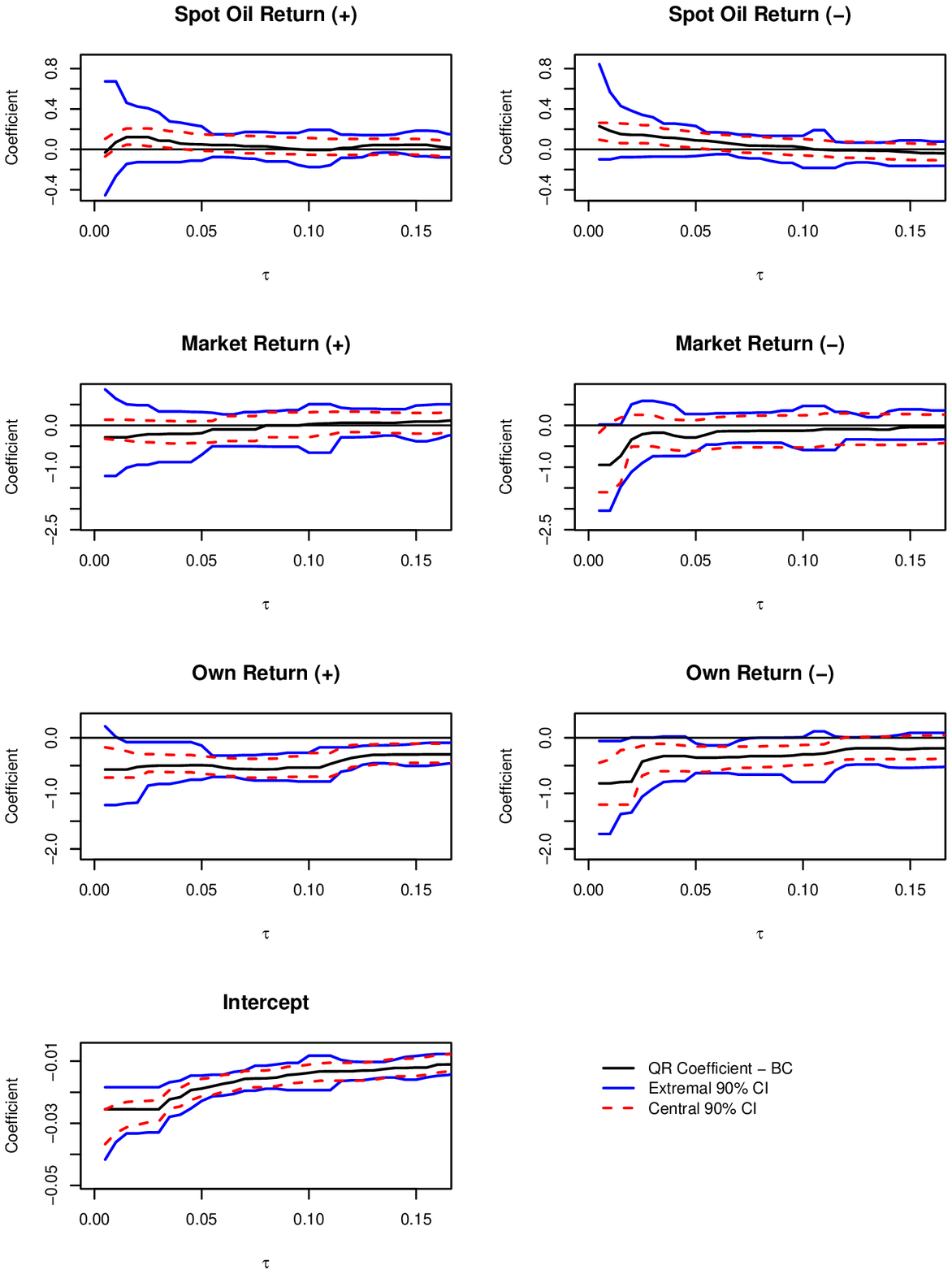, width=6in,height=8in}
\caption{Bias-corrected QR coefficient estimates and $90\%$
pointwise confidence intervals for $\tau \leq .15$. The solid lines
depict extremal confidence intervals. The dashed lines depict normal
confidence intervals. }
 \end{center}
\end{figure}

\begin{figure}[htbp!]\label{result4}
\begin{center}
\noindent
 \centering\epsfig{figure=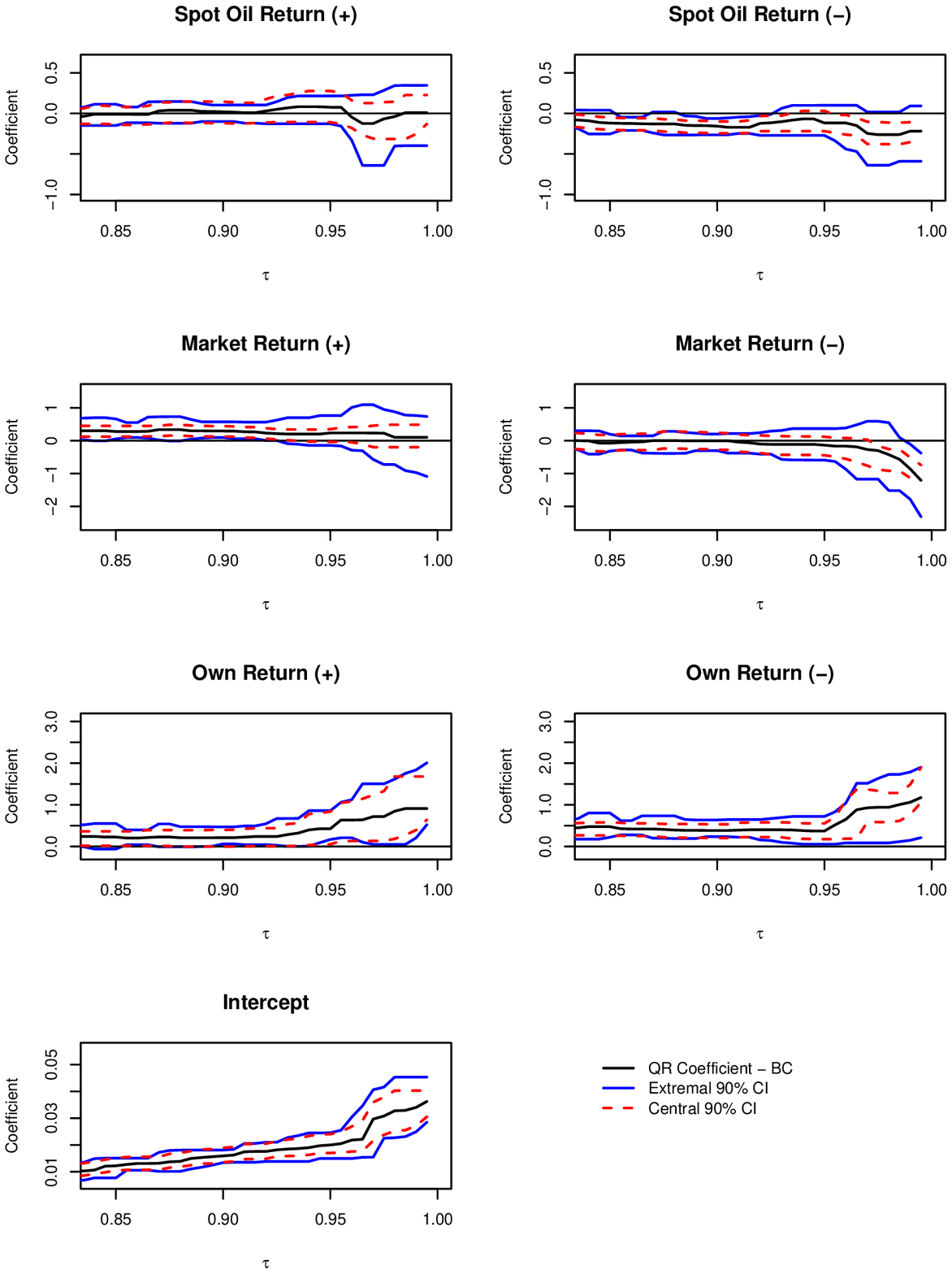, width=6in,height=8in}
\caption{Bias-corrected QR coefficient estimates and $90\%$
pointwise intervals for $\tau \geq .85$ .  The solid lines depict
extremal confidence intervals. The dashed lines depict normal
confidence intervals. }
\end{center}
\end{figure}


\begin{figure}[htbp!]\label{densities}
\begin{center}
\noindent
 \centering\epsfig{figure=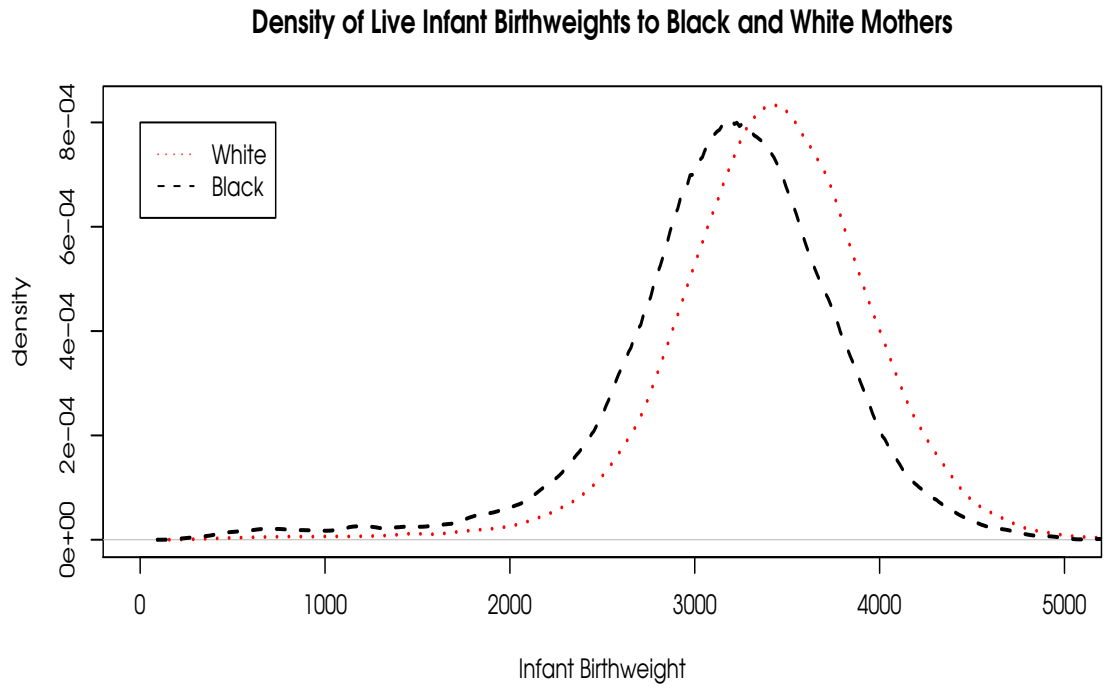, width=6in,height=4in}
\caption{Birthweight densities for black and white mothers.}
\end{center}
\end{figure}

\begin{figure}[htbp!]\label{result4}
\begin{center}
\noindent
 \centering\epsfig{figure=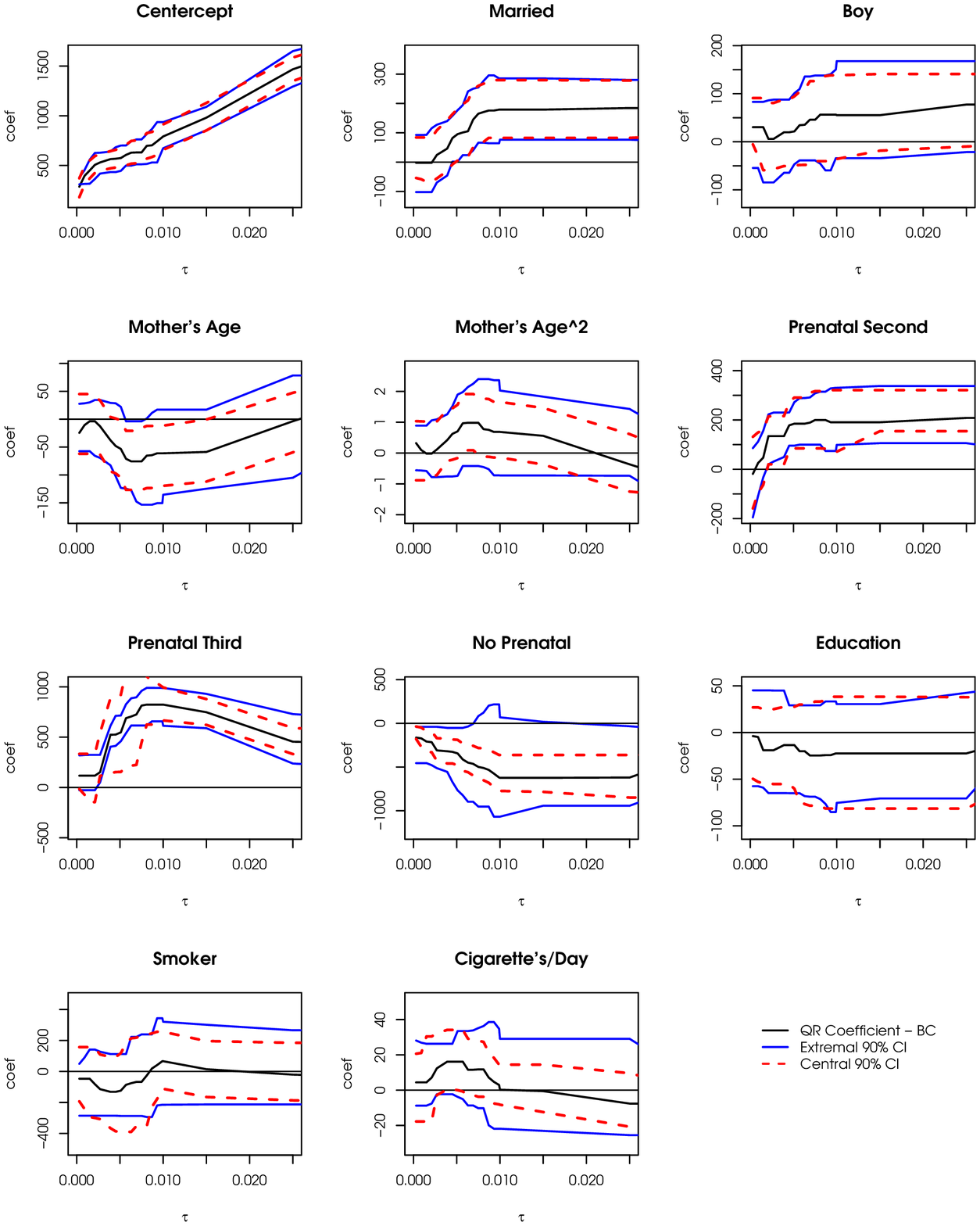, width=6in,height=8in}
\caption{Bias-corrected QR coefficient estimates and $90\%$
pointwise confidence intervals for $\tau \leq .025$ .  The solid
lines depict extremal confidence intervals. The dashed lines depict
normal confidence intervals. }
\end{center}
\end{figure}

\begin{figure}[htbp!]\label{result4}
\begin{center}
\noindent
 \centering\epsfig{figure=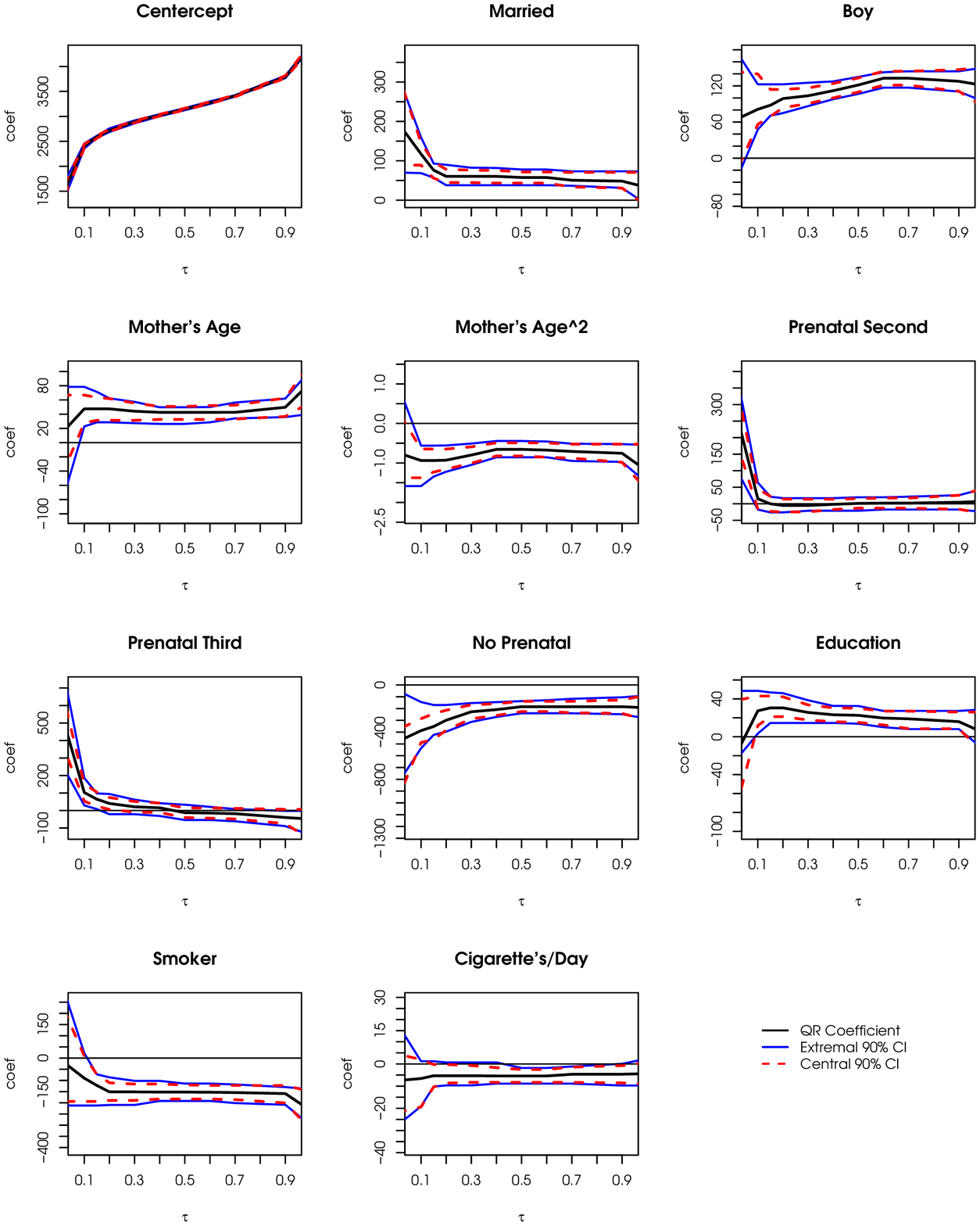, width=6in,height=8in}
\caption{QR coefficient estimates and $90\%$ pointwise confidence
intervals for $\tau \in [.025, .975]$ . The solid lines depict
extremal confidence intervals. The dashed lines depict normal
confidence intervals. }
\end{center}
\end{figure}

\end{document}